\documentclass[twocolumn]{aastex63}

\usepackage{natbib}
\usepackage{hyperref}
\usepackage{floatrow}
\usepackage{amsmath,amssymb}
\usepackage{ulem}
\shorttitle{Star formation in GMF G45.3+0.1}
\shortauthors{N.~K. Bhadari et al.}

\begin{document}

\title{Simultaneous evidence of edge collapse and hub-filament configurations: A rare case study of a Giant Molecular Filament G45.3+0.1}

\correspondingauthor{N.~K. Bhadari}
\email{Email: naval@prl.res.in}

\author{N.~K. Bhadari}
\affiliation{Astronomy \& Astrophysics Division, Physical Research Laboratory, Navrangpura, Ahmedabad 380009, India}
\affiliation{Indian Institute of Technology Gandhinagar Palaj, Gandhinagar 382355, India.}

\author{L.~K. Dewangan}
\affil{Astronomy \& Astrophysics Division, Physical Research Laboratory, Navrangpura, Ahmedabad 380009, India}

\author{D.~K. Ojha}
\affil{Department of Astronomy and Astrophysics, Tata Institute of Fundamental Research, Homi Bhabha Road, Mumbai 400005, India.}

\author{L.~E. Pirogov}
\affil{Institute of Applied Physics of the Russian Academy of Sciences, 46 Ulyanov st., Nizhny Novgorod 603950, Russia.}

\author{A.~K. Maity}
\affiliation{Astronomy \& Astrophysics Division, Physical Research Laboratory, Navrangpura, Ahmedabad 380009, India}
\affiliation{Indian Institute of Technology Gandhinagar Palaj, Gandhinagar 382355, India.}


\begin{abstract}
We study multiwavelength and multiscale data to investigate the kinematics of molecular gas associated with the star-forming complexes G045.49+00.04 (G45E) and G045.14+00.14 (G45W) in the Aquila constellation.
An analysis of the FUGIN $^{13}$CO(1--0) line data unveils the presence of a giant molecular filament (GMF G45.3+0.1; length $\sim$75 pc, mass $\sim$1.1$\times$10$^{6}$ M$_{\odot}$) having a coherent velocity structure at [53, 63] km s$^{-1}$.
The GMF G45.3+0.1 hosts G45E and G45W complexes at its opposite ends.
We find large scale velocity oscillations along GMF G45.3+0.1, which also reveals the linear velocity gradients of $-$0.064 and $+$0.032 km s$^{-1}$ pc$^{-1}$ at its edges.
The photometric analysis of point-like sources shows the clustering of young stellar object (YSO) candidate sources at the filament's edges where the presence of dense gas and H{\sc ii} regions are also spatially observed.
The {\it Herschel} continuum maps along with the CHIMPS $^{13}$CO(3--2) line data unravel the presence of parsec scale hub-filament systems (HFSs) in both the sites, G45E and G45W.
Our study suggests that the global collapse of GMF G45.3+0.1 is end-dominated, with addition to the signature of global nonisotropic collapse (GNIC) at the edges.
Overall, GMF G45.3+0.1 is the first observational sample of filament where the edge collapse and the hub-filament configurations are simultaneously investigated. These observations open up the new possibility of massive star formation, including the formation of HFSs.
\end{abstract}



\section{Introduction}
\label{sec:intro}

Based on the analysis of infrared and submillimeter continuum maps, networks of filaments or hub-filament systems \citep[HFSs;][]{Myers09} have been recognized as potential birthplaces of massive stars and clusters of young protostars \citep[e.g.,][]{Andre10,Schneider2012,Anderson2021}, and such HFSs are commonly detected structures in our Galaxy \citep[e.g.,][see references therein]{Motte2018}.
In general, the HFS is a junction of multiple filaments. 
In such configurations, filaments can be depicted with high aspect ratio (length/diameter), and the central hub is identified with low aspect ratio \citep[e.g.,][]{Myers09}. Molecular line observations unveil that the filaments in HFSs act as a channel of material flows to feed the dense clumps/cores \citep[e.g.,][]{Kirk2013,Peretto2013,Dewangan2017b,Hacar2018,Morales2019,Chen2020ApJ,Dewangan2020,Dewangan2021MNRAS,dewangan22}, where massive stars and young stellar clusters are observed \citep[][and references therein]{Liu2012,Morales2019}.
In addition to the HFSs, isolated filaments associated with star-forming activities have also been identified in our Galaxy.
In particular, concerning isolated filaments, the onset of the end-dominated collapse (EDC; or edge collapse) process has been reported in a very few star-forming sites \citep[including H{\sc ii} regions powered by massive stars;][and references therein]{Dewangan2017,Dewangan19S242,Bhadari20}, where one exclusively expects two clumps/cores produced at the respective end of filament via high gas acceleration \citep[e.g.,][]{Bastien83,Pon12,Clarke15,Hoemann2022arXiv}. 
These observed configurations directly manifest the role of filaments in star formation processes. However, no attempt has been made to explore the connection between these two distinct filament configurations. In this context, the present paper deals with two major star-forming complexes (i.e., G045.49+00.04 and G045.14+00.14) located in the Aquila constellation ($l$=45$^{o}$, $b$=0$^{o}$). These sites were primarily identified and catalogued as extended clouds with the prefix name of Galactic Ring Survey Molecular Cloud \citep[GRSMC; see more details in][]{Rathborne09}.

\begin{figure*}
\includegraphics[width=0.8\textwidth]{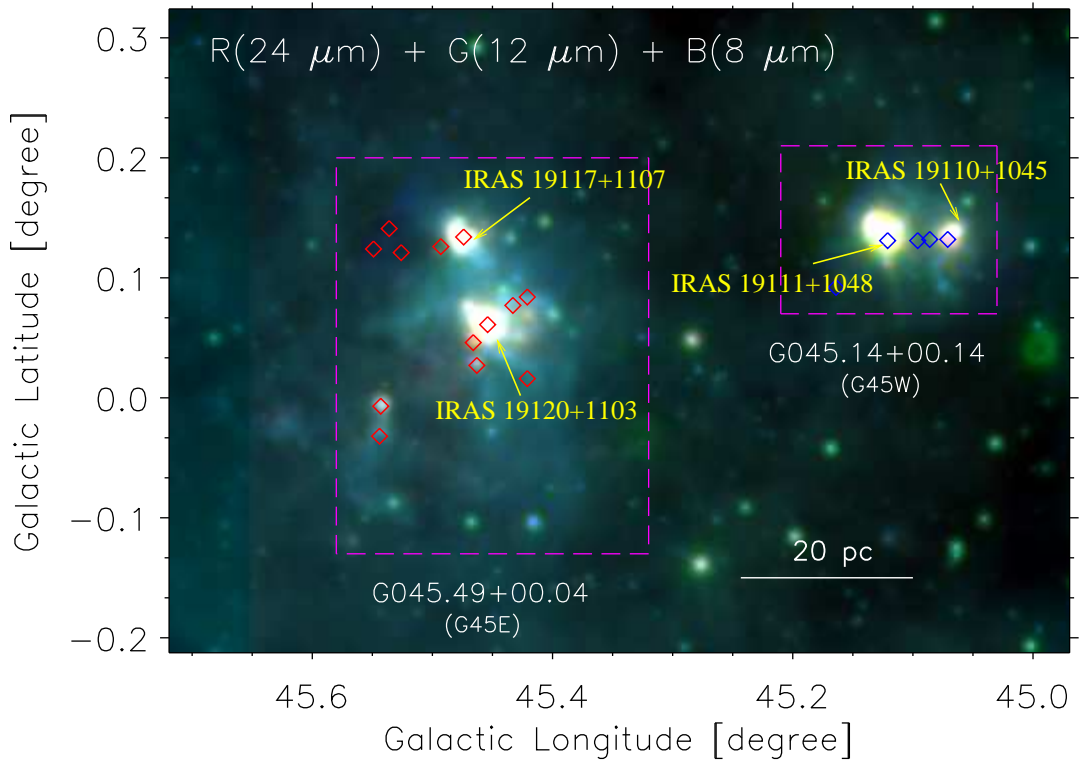}
\caption{Three-color composite image (Red: {\it Spitzer} 24 $\mu$m, Green: {\it WISE} 12 $\mu$m, Blue: {\it Spitzer} 8 $\mu$m) of a field containing star-forming complexes G045.49+00.04 (G45E) and G045.14+00.14 (G45W).
Two dashed rectangular boxes (in magenta) enclose the areas of G45E and G45W \citep[from GRSMC;][]{Rathborne09}, and are presented in Figures~\ref{fig2}b and \ref{fig2}c.
The red and blue diamonds represent the positions of ATLASGAL dust clumps at distances of 8.4 kpc and 8.0 kpc, respectively, from \citet{Urquhart18}.
The positions of the IRAS sources are shown by arrows.
A scale bar of 20 pc (at $d$ = 8 kpc) is also displayed.
}
\label{fig1}
\end{figure*}

The two complexes are well known for hosting massive star-forming regions, numerous ultracompact (UC) H{\sc ii} regions, maser emission, and outflow activities from young stars including massive star(s) \citep[e.g.,][]{Wood89, Blum08,Paron09, Rivera-Ingraham10}. 
The eastern complex G045.49+00.04 (hereafter, G45E) harbors the sources IRAS 19120+1103 (or G45.45+0.06) 
and IRAS 19117+1107 (or G45.48+0.13), while the western complex G045.14+00.14 (hereafter, G45W) hosts 
two active massive star-forming regions (IRAS 19110+1045 and IRAS 19111+1048). Both the complexes have been well explored 
using the infrared, submillimeter, and radio data sets \citep[e.g.,][]{Vig06,Rivera-Ingraham10}. 
In particular, the eastern complex is associated with the significant extended submillimeter emission \citep[see Figure~2 in][]{Rivera-Ingraham10}. 
In the direction of the western complex, the distribution of dust and ionized gas around IRAS 19110+1045 is less extended than that of IRAS 19111+1048 \citep{Vig06}, and the former source is observed to be much younger than the other one \citep{Hunter97,Vig06}.
A mid-infrared (MIR) view of an area containing G45E and G45W is presented in Figure~\ref{fig1}, which is a 3-color composite map made using the {\it Spitzer} 24 $\mu$m (red), {\it WISE} 12 $\mu$m (green), and {\it Spitzer} 8 $\mu$m (blue) images. 
 
Figure~\ref{fig1} is also overlaid with the positions of the APEX Telescope Large Area Survey of the Galaxy \citep[ATLASGAL;][]{Schuller09} dust continuum clumps at 870 $\mu$m  identified by \citet{Urquhart18}. Radial velocity and distance information of each ATLASGAL clump are available \citep[see][for more details]{Urquhart18}. In Figure~\ref{fig1}, the ATLASGAL 870 $\mu$m dust clumps are marked by red diamonds and blue diamonds, and are traced in a velocity range of [55.5, 62.2] km s$^{-1}$.
The clumps shown with red diamonds and blue diamonds are located at the kinematic distances of 8.4 kpc and 8 kpc, respectively. 
However, based on the spatial distribution of the ATLASGAL clumps, and the morphology of the parent molecular cloud toward the sites G45E and G45W (Section~\ref{kinematics_g45}), we adopt a distance of 8~kpc for both the sites in this paper. Previously, distances to these sites were considered in the range of $\sim$6--8 kpc \citep[see][and references therein]{Rivera-Ingraham10}.
A careful study of the molecular gas containing both the complexes is yet to be carried out. Such study is essential to probe the ongoing physical processes and will also allow us to explore the HFSs and EDC scenarios. In this relation, the submillimeter continuum maps are also utilized to carefully examine the observed spatial structures. 

We organize this paper into five major sections.
Section~\ref{sec:obser} describes data sets used in this paper.
In Section~\ref{sec:results}, we present the results obtained using the continuum and line data. 
In Section~\ref{sec:discussion}, we discuss the implications of our derived results for explaining the star formation processes. 
A summary of this work is given in Section~\ref{sec:conc}.
\section{Data Sets}
\label{sec:obser}
%


In this work, we analyzed the publically available multiscale and multiwavelength data sets toward our target region, primarily in the area of 0$\degr$.67$\times$0$\degr$.50 or 93.5$\times$70.4 pc$^{2}$ (centred at $l$ = 45$\degr$.30; $b$ = 0$\degr$.05). The overview of data sets used in this work is presented in Table~\ref{tab1}.
We utilized the $^{12}$CO(J=1$-$0), $^{13}$CO(J=1$-$0), and C$^{18}$O(J=1$-$0) line data obtained from the FOREST Unbiased Galactic plane Imaging survey with the Nobeyama 45-m telescope (FUGIN) survey. The FUGIN data are calibrated in main beam temperature ($T_{\rm mb}$), and the typical rms is $\sim$1.5~K for $^{12}$CO and $\sim$0.7~K for $^{13}$CO and C$^{18}$O lines. The survey data provide a velocity resolution of 1.3 km s$^{-1}$ \citep[see more details in][]{umemoto17}.
We also used $^{13}$CO(J=3$-$2), and C$^{18}$O(J=3$-$2) line data downloaded from the $^{13}$CO/C$^{18}$O (J = 3$\rightarrow$2) Heterodyne Inner Milky Way Plane Survey (CHIMPS).
The high-resolution ($\theta_{\rm spatial}$=15$''$, $\theta_{\rm velocity}$=0.5 km s$^{-1}$) CHIMPS survey covers the Galactic plane within $\lvert b \rvert$ $\leq$ 0$\degr$.5 and 28$\degr$ $\lesssim$ $l$ $\lesssim$ 46$\degr$. The survey data have a median rms of $\sim$0.6 K, and are calibrated in antenna temperature ($T_{\rm A}^{*}$) scale. We converted the intensity scales from $T_{\rm A}^{*}$ to $T_{\rm mb}$ by using the relation $T_{\rm mb}=\frac{T_{\rm A}^{*}}{\eta_{\rm mb}}$, where the mean detector efficiency is $\eta_{\rm mb}$= 0.72 \citep{Rigby16}.
In order to improve the sensitivity, the molecular line data from the FUGIN and CHIMPS were smoothed with a Gaussian function having 3 pixels half-power beamwidth. This exercise yields the final spatial resolutions of the FUGIN and CHIMPS molecular line data to be 
21\rlap.{$''$}73 and 16\rlap.{$''$}82, respectively.

 \begin{table*}
\scriptsize
\setlength{\tabcolsep}{0.025in}
\centering
\caption{List of multiwavelength surveys used in this paper}
\label{tab1}
\begin{tabular}{lcccr}
\hline 
  Survey  &  Wavelength/line(s)      &  Angular Resolution ($''$ )        &  Reference \\   
\hline

Multi-Array Galactic Plane Imaging Survey (MAGPIS) &20 cm	& $\sim$6	&\citet{Helfand06}\\

FUGIN
& $^{12}$CO, $^{13}$CO, C$^{18}$O (J=1--0) & $\sim$20       &\citet{umemoto17}\\

CHIMPS
&$^{13}$CO, C$^{18}$O (J=3--2)	&$\sim$15	&\citet{Rigby16}\\

APEX Telescope Large Area Survey of the Galaxy (ATLASGAL) & 870 $\mu$m & $\sim$19.2 &\citet{Schuller09}\\

{\it Herschel} Infrared Galactic Plane Survey (Hi-GAL)       &70, 160, 250, 350, 500 $\mu$m                     &$\sim$6, 12, 18, 25, 37   &\citet{Molinari10}\\

Spitzer MIPS Inner Galactic Plane Survey (MIPSGAL)	&24 $\mu$m 	&$\sim$6	&\citet{Carey05}\\ 

Wide Field Infrared Survey Explorer (WISE) & 12 $\mu$m            & $\sim$6.5           &\citet{Wright10}\\ 

$\it Spitzer$ Galactic Legacy Infrared Mid-Plane Survey Extraordinaire (GLIMPSE)      &3.6, 4.5, 5.8, 8.0 $\mu$m                   & $\sim$2          &\citet{Benjamin03}\\

UKIRT near-infrared Galactic Plane Survey (GPS)            &1.25--2.2 $\mu$m                   &$\sim$0.8           &\citet{Lawrence07}\\ 

Two Micron All Sky Survey (2MASS)              &1.25--2.2 $\mu$m                  & $\sim$2.5          &\citet{Skrutskie06}\\
\hline          
\end{tabular}
\end{table*}

\section{Results}
\label{sec:results}
\subsection{Kinematics of molecular gas}
\label{kinematics_g45}

In this section, we study the spatial-kinematic structures of the molecular gas in the direction of our target sites.

\subsubsection{Morphology of the molecular cloud}

\begin{figure*}
\includegraphics[width=0.8\textwidth]{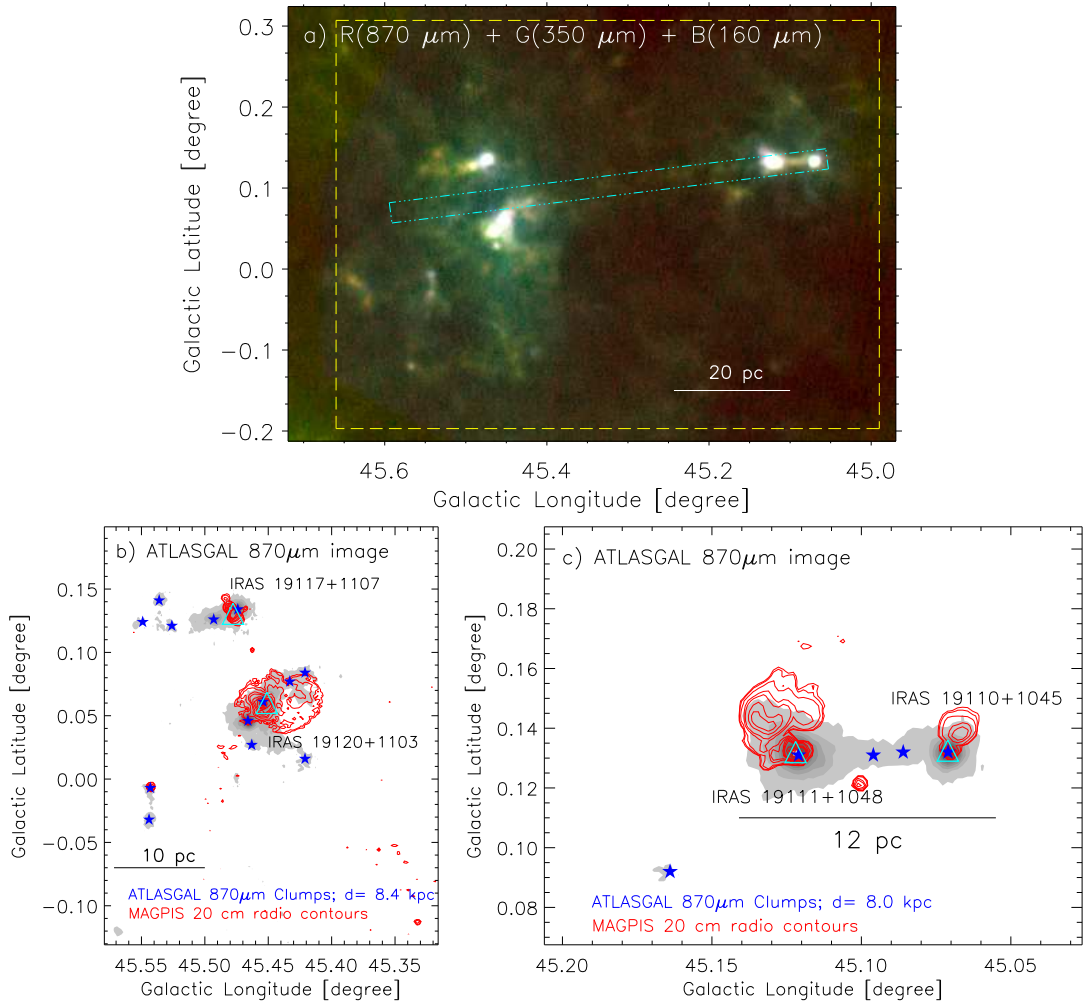}
\caption{(a) Three-color composite image (Red: ATLASGAL 870 $\mu$m, Green: {\it Herschel} 350 $\mu$m, Blue: {\it Herschel} 160 $\mu$m) of a field shown in Figure~\ref{fig1}.
A dotted dashed rectangular strip (in spring green) indicates an area where the averaged gas spectra are extracted (see Figure~\ref{fig3}). A dashed box in yellow highlights an area studied in this work.
Panels (b) and (c) display the ATLASGAL dust continuum contour maps (in grey scale) of G045.49+00.04 and G045.14+00.14 regions at 870 $\mu$m, respectively (see dashed boxes in Figure~\ref{fig1}). In panels (b) and (c), the ATLASGAL contour levels range from 5\% to 95\% in steps of 9\%, of the map peak value (i.e., 5.49(7.56) Jy beam$^{-1}$ for panel ``b"(``c")).
The maps are overlaid with the MAGPIS 20 cm continuum emission contours with the levels of 0.96, 1.36, 2.73, 8.19, 13.64, 27.29, 40.93, 54.57, 109.15, 163.72, 218.30, 259.23, 267.41 mJy beam$^{-1}$. The star symbols in panels (b) and (c) represent the positions of ATLASGAL dust clumps (see also Figure~\ref{fig1}), and the cyan triangles indicate the positions of IRAS sources. 
}
\label{fig2}
\end{figure*}

In Figure~\ref{fig2}a, we display a three-color composite map (Red: ATLASGAL 870 $\mu$m, Green: {\it Herschel} 350 $\mu$m, Blue: {\it Herschel} 160 $\mu$m) at submillimeter wavelengths of the target area presented in Figure~\ref{fig1}.
The map unravels the presence of elongated structures (or filaments) in both the star-forming complexes, G45E and G45W.
Figures~\ref{fig2}b and \ref{fig2}c show the ATLASGAL 870 $\mu$m continuum images overlaid with the MAGPIS 20 cm radio contours for the sites G45E and G45W, respectively. The MAGPIS radio contours display the zones of ionized emission or the H{\sc ii} regions in the target sites.

\begin{figure*}
\includegraphics[width=0.6\textwidth]{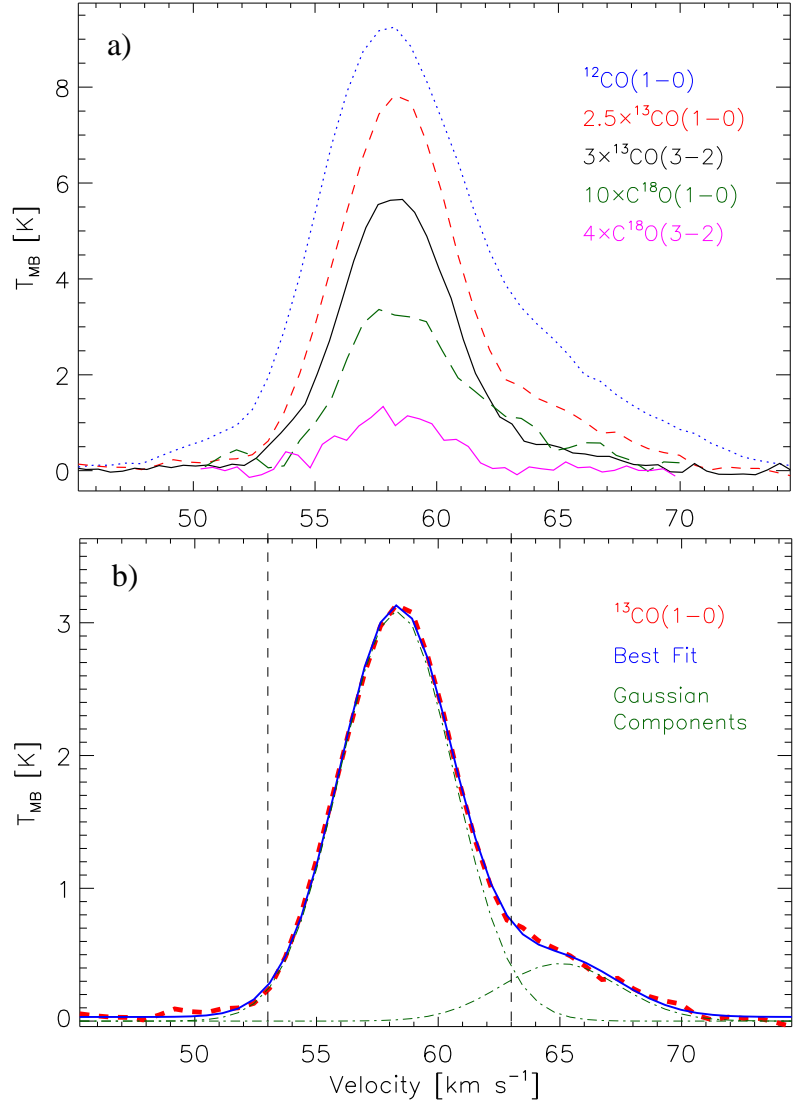}
\caption{(a) Averaged spectral profiles of different molecular species along the dotted-dashed strip shown in Figure~\ref{fig2}a.
(b) $^{13}$CO(1--0) line profile (dashed (red) curve) fitted with the two Gaussian components (dotted-dashed (green) curves).
The best fit profile is shown in a thick blue curve.
The two vertical dashed lines are marked at the velocity of 53 and 63 km s$^{-1}$.
}
\label{fig3}
\end{figure*}

In order to study the kinematics of the parent molecular cloud, we have extracted the averaged molecular spectra toward a rectangular strip that passes through both the major star-forming complexes (see dotted dashed rectangle in Figure~\ref{fig2}a). Figure~\ref{fig3}a presents the averaged spectra of different molecular species (i.e., FUGIN $^{12}$CO/$^{13}$CO/C$^{18}$O (1--0), CHIMPS $^{13}$CO/C$^{18}$O (3--2)).
The various line profiles suggest that the molecular cloud can be studied in a velocity range of [50, 70] km s$^{-1}$. 
The $^{12}$CO(1--0) profile appears much broader than the other line profiles, which can be explained due to 
the high optical depth and lower critical density of $^{12}$CO(1--0). 
In general, the $^{12}$CO(1--0) line data are known to only trace the outer diffuse cloud. 
Hence, we considered a relatively optically thin $^{13}$CO(1--0) line data to depict the molecular cloud boundary. 
The {\texttt curve\_fit} function from the {\texttt SciPy} library \citep{scipy20} is employed to fit the $^{13}$CO(1--0) spectra with double Gaussian profiles.
The fitted Gaussian profiles along with the original $^{13}$CO(1--0) profile are shown in Figure~\ref{fig3}b.
The major Gaussian component is traced in a velocity range of [53, 63] km s$^{-1}$, and the fitting procedure gives the mean ($\mu$) and standard deviation ($\sigma$) to be 58.30$\pm$0.03 km s$^{-1}$ and 2.34$\pm$0.03 km s$^{-1}$, respectively.
The other Gaussian component is associated with $\mu$= 65.01$\pm$0.20 km s$^{-1}$ and $\sigma$= 2.43$\pm$0.19 km s$^{-1}$, and contributes to the redder tail of the observed $^{13}$CO(1--0) profile.

\begin{figure*}
\includegraphics[width=\textwidth]{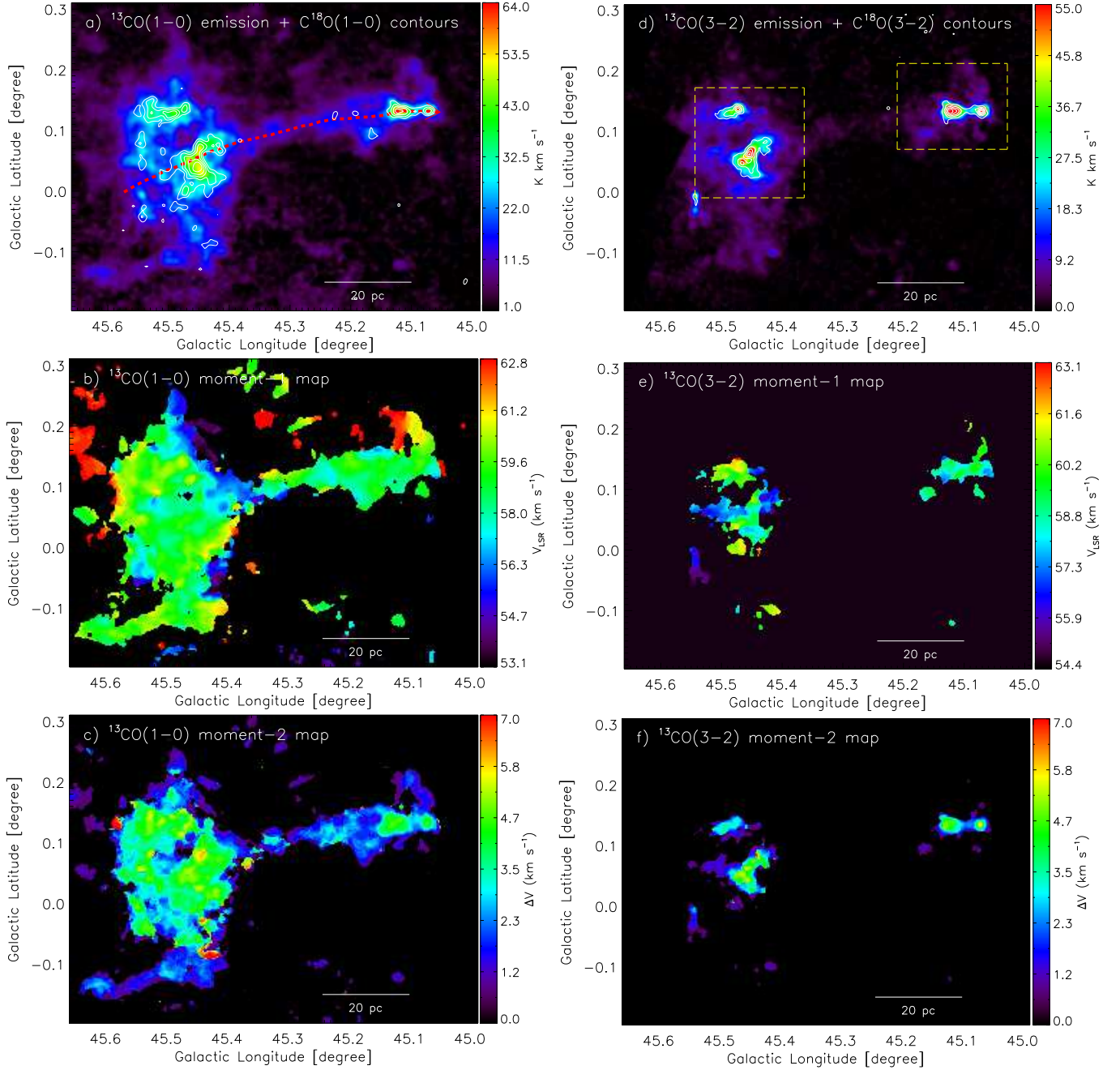}
\caption{Left column of panels shows the FUGIN $^{13}$CO(1--0) moment maps; (a) moment-0 map, (b) moment-1 map, and (c) moment-2 map in the direction of our selected target area (see a dashed box in Figure~\ref{fig2}a). Right column of panels shows the similar maps for CHIMPS $^{13}$CO(3--2) emission. 
The moment-0 maps are obtained in the velocity range of [53, 63] km s$^{-1}$.
In panel (a), C$^{18}$O(1--0) emission contours are overlaid on the  $^{13}$CO(1--0) moment-0 map. The contour levels range from 3 to 8.7 K km s$^{-1}$ in steps of 1.3 K km s$^{-1}$.
In panel (d), the C$^{18}$O(3--2) emission contours are overlaid with the levels ranging from 3.2 to 20 K km s$^{-1}$ in steps of 3.36 K km s$^{-1}$.
Two dashed boxes in panel (d) show the area presented in Figure~\ref{fig12}.
In panel (a), a dashed (red) curve is indicative of filament's length, where the position-velocity diagram is also studied (see Figure~\ref{fig6}c).
In all the panels, the respective color bars are shown to the right of each panel.
}
\label{fig4}
\end{figure*}

In Figure~\ref{fig4}, we have shown the spatial distribution of the molecular cloud at [53, 63] km s$^{-1}$ 
associated with the major Gaussian component. 
Figure~\ref{fig4}a displays the integrated intensity (moment-0) map of $^{13}$CO(1--0) emission, which is also overlaid with the C$^{18}$O (1--0) emission contours.
The $^{13}$CO(1--0) moment-0 map reveals the presence of an elongated molecular cloud that hosts the two previously star-forming complexes, G45E and G45W, at its ends.
The C$^{18}$O (1--0) emission further unveils the dense substructures in the direction of major massive star-forming sites (i.e., IRAS 19120+1103, IRAS 19117+1107, IRAS 19110+1045 and IRAS 19111+1048).

The intensity weighted mean velocity (moment-1) map of the $^{13}$CO(1--0) emission is shown in Figure~\ref{fig4}b. A continuous velocity structure around 58 km s$^{-1}$ further supports the idea of an elongated molecular cloud. 
We can also notice a relatively redshifted emission around 63 km s$^{-1}$, which belongs to the minor Gaussian component (Figure~\ref{fig3}). However, the moment maps related to the minor Gaussian component are not studied in this paper. 
The intensity weighted linewidth map, which is generally known as the moment-2 map, is shown in Figure~\ref{fig4}c. The linewidth value is about $\sim$2--5 km s$^{-1}$ throughout the cloud structure.
There are clear velocity dispersion enhancements at the edges of the cloud, which are of the order of $\sim$4.5 km s$^{-1}$.
We have also displayed the moment maps of $^{13}$CO(3--2) emission in Figures~\ref{fig4}d, \ref{fig4}e, and \ref{fig4}f.
The $^{13}$CO(3--2) emission traces the dense gas compared to the $^{13}$CO(1--0) emission, thus allowing us to study the structures and kinematics of dense molecular clumps. From Figures~\ref{fig4}a and \ref{fig4}d, the dense gas/clumps are primarily identified towards the ends of elongated cloud, and the entire cloud structure is traced in the $^{13}$CO(1--0) emission only.
Overall, Figure~\ref{fig4} collectively displays the spatial-kinematic distribution of molecular cloud associated with the target sites G45E and G45W.

\begin{figure*}
\includegraphics[width=\textwidth]{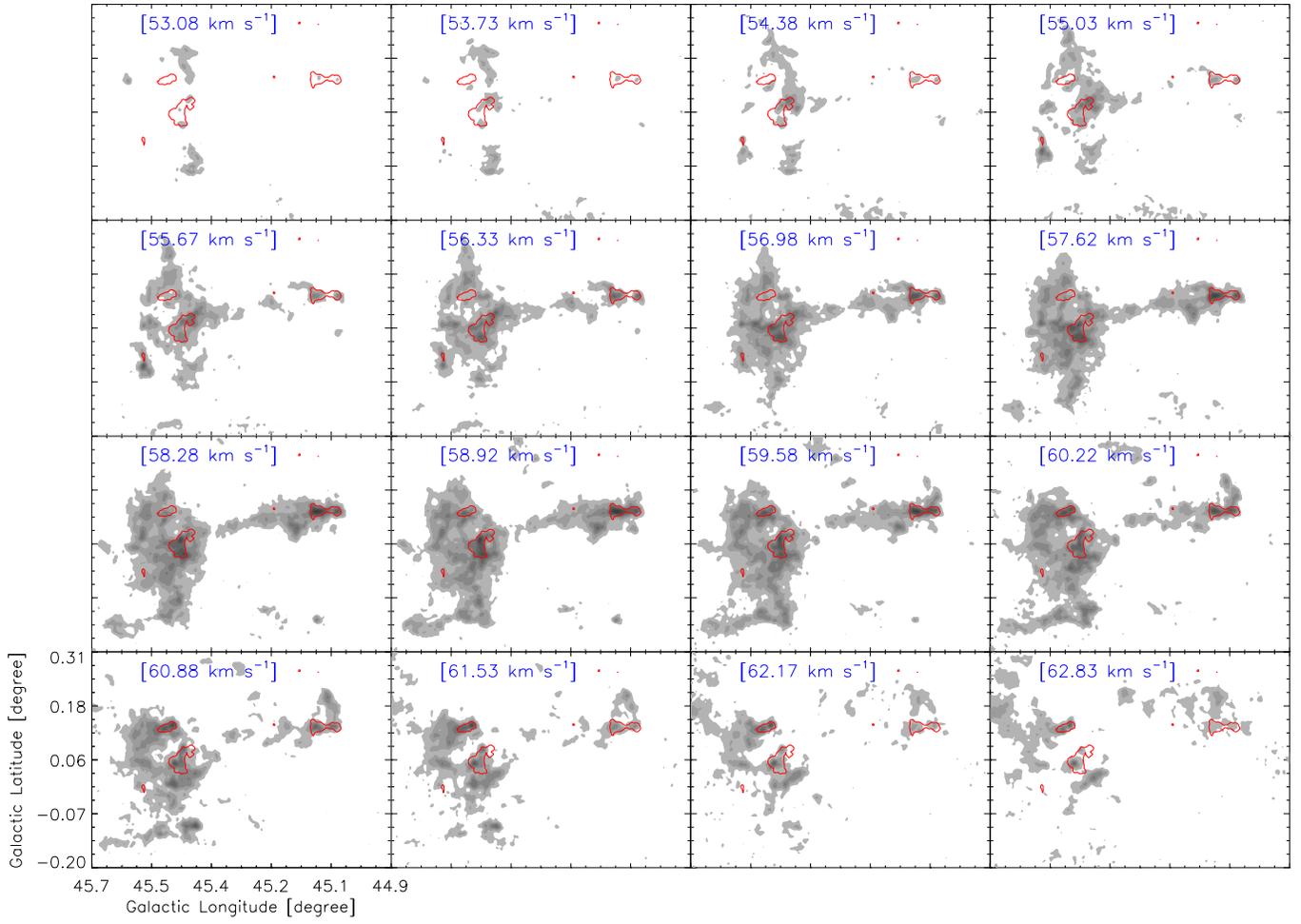}
\caption{Velocity channel maps of $^{13}$CO(1--0) emission (in grayscale). The contour levels range from 1.2 to 14 K in steps of 1.28 K. 
The overlaid red contours represent the integrated emission of C$^{18}$O(3--2) at a level of 3.2 K km s$^{-1}$ (see also Figure~\ref{fig4}b).
The velocities are labeled in each panel.
}
\label{fig5}
\end{figure*}

To examine the velocity distribution of the gas toward the elongated cloud observed in a velocity range of [53, 63] km s$^{-1}$, we have shown the velocity channel maps in Figure~\ref{fig5}. To infer the gas motion in the direction of the cloud, in Figure~\ref{fig5}, we have also shown the C$^{18}$O(3--2) integrated emission contour (in red) at a level of 3.2 K km s$^{-1}$. The elongated cloud morphology starts to appear from 55.67 km s$^{-1}$ and takes a complete shape in the velocity range from 57.62 km s$^{-1}$ to 59.58 km s$^{-1}$. The dense substructures are mostly seen at the edges of the elongated cloud.

\subsubsection{Position-velocity diagrams}
\label{pvdiagrams}

\begin{figure*}
\includegraphics[width=0.8\textwidth]{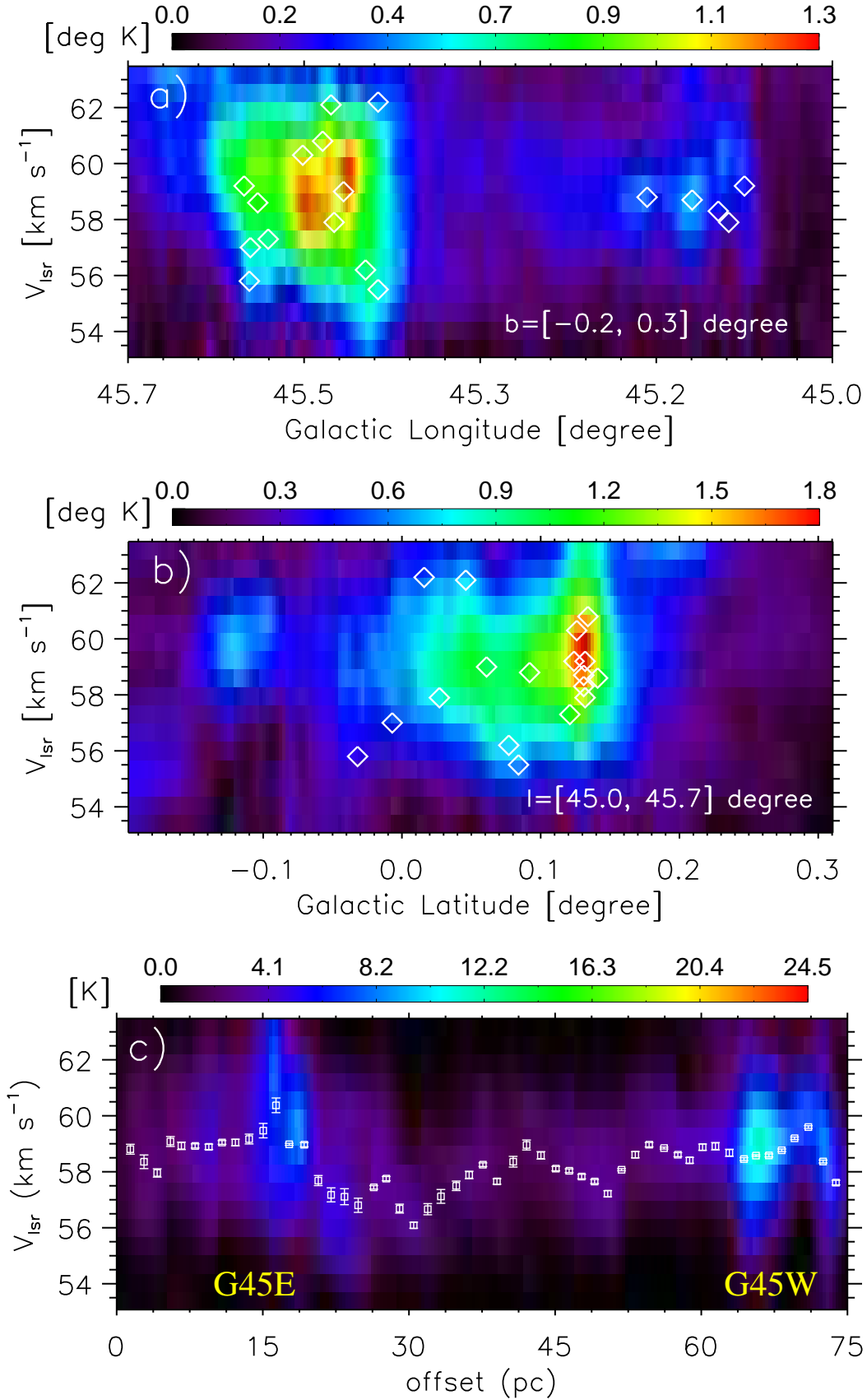}
\caption{Position-velocity (pv) diagrams of $^{13}$CO(1--0) emission: (a) Longitude-velocity diagram, (b) Latitude-velocity diagram, toward an area displayed in Figure~\ref{fig5}.
The diamond symbols represent the positions of ATLASGAL clumps from \citet{Urquhart18} (see also Figure~\ref{fig1}). 
(c) The pv diagram along a dashed curve marked in Figure~\ref{fig4}a, where zero point in the x-axis corresponds to the eastern end of the curve. The data points with error bars are the velocity averaged points at their respective position (see text for more details).
}
\label{fig6}
\end{figure*}

\begin{figure*}
\includegraphics[width=0.8\textwidth]{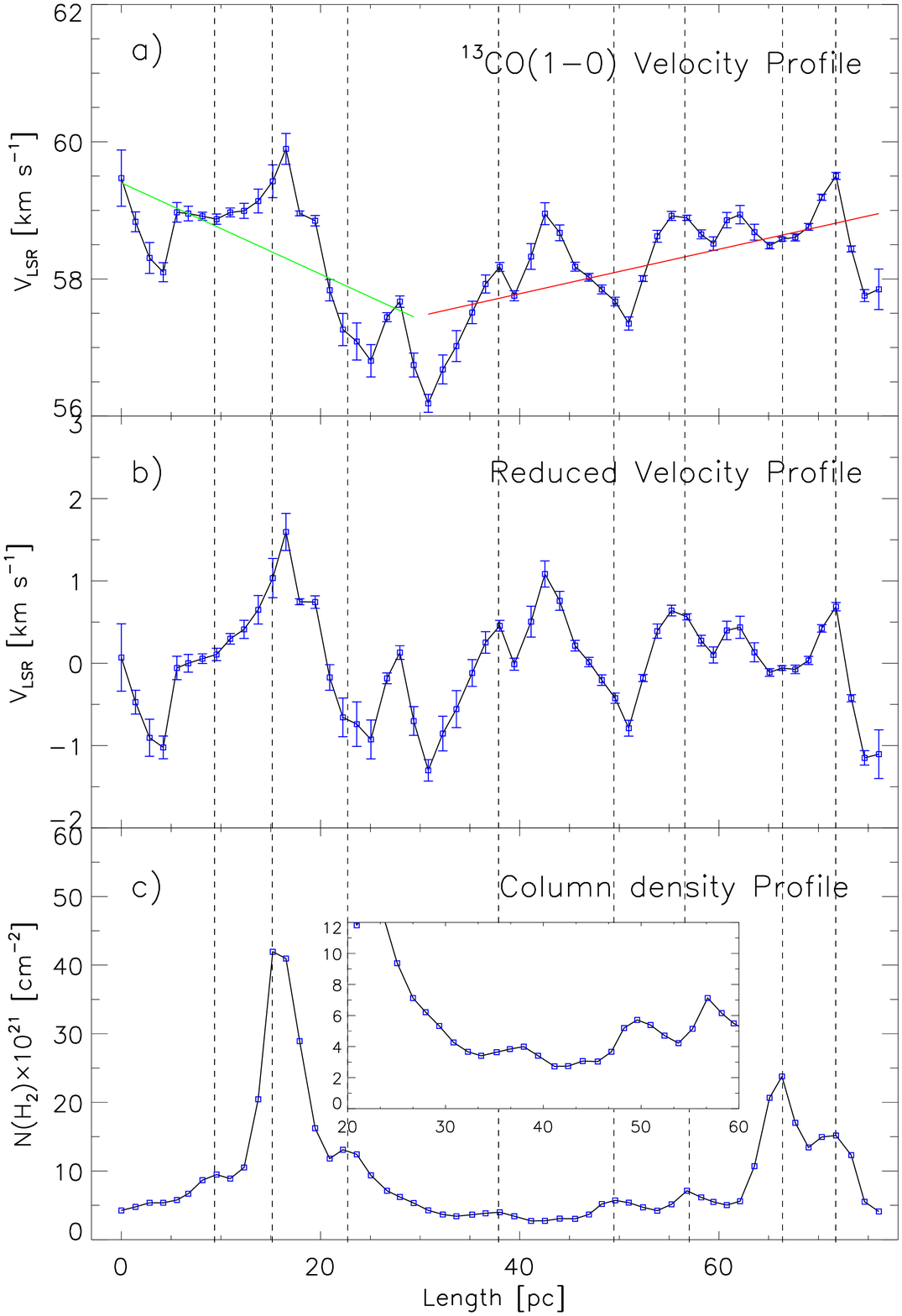}
\caption{(a) Position-velocity profile of $^{13}$CO(1--0) emission, similar to that shown in Figure~\ref{fig6}c. The green and red lines indicate the velocity gradient of $-$0.064 km s$^{-1}$ pc$^{-1}$ and $+$0.032 km s$^{-1}$ pc$^{-1}$, respectively.
(b) Residual velocity profile after removing the velocity gradients shown in panel (a).
(c) Column density profile. The inset shows a similar column density profile in the length range of 20 to 60 pc.
}
\label{fig7}
\end{figure*}

We have also examined the position-velocity (pv) diagrams to infer the kinematics of molecular gas. A pv diagram is an important tool to investigate the gas motion along any spatial axis.
The longitude-velocity ({\it l--v}) and latitude-velocity ({\it b--v}) diagrams are displayed in Figures~\ref{fig6}a and \ref{fig6}b, respectively. We have also overlaid the positions of ATLASGAL dust clumps on the {\it l--v} and {\it b--v} diagrams, showing the spatial and velocity association of the dust clumps with the cloud. 
A pv diagram along a selected path (physical length $\sim$75 pc) is shown in Figure~\ref{fig6}c. This path is highlighted by a dashed (red) curve in Figure~\ref{fig4}a. Several small circular regions (radii $\sim$ 15$''$) along this path are also selected, 
and an average $^{13}$CO(1--0) profile over each circular region is produced. For each average profile, the peak velocity is derived 
through the fitting of a Gaussian function. In Figure~\ref{fig6}c, to study the pattern of velocity, we have also overlaid the peak velocity of each observed $^{13}$CO(1--0) profile. 
In the pv diagram, the velocity distribution along the path shows an oscillatory pattern, and has a maximum velocity variation of $\sim$4 km s$^{-1}$.
In addition to the oscillatory pattern, the velocity profile shows the linear gradient at both ends of the filament.
In the direction of G45E, the velocity gradient is observed to have a value of $-$0.064 km s$^{-1}$ pc$^{-1}$, while it is $+$0.032 km s$^{-1}$ pc$^{-1}$ toward G45W.
Figure~\ref{fig7}a highlights two straight lines, indicating the presence of different velocity gradients in the pv plot (see also Figure~\ref{fig6}c).
We also note that the largest velocity gradient ($\sim$0.5 km s$^{-1}$ pc$^{-1}$) is associated with the western border of G45E. This signature could be an indication of accelerated gas motions.
The residual velocity profile is shown in Figure~\ref{fig7}b, which is generated after removing the mentioned velocity gradients.
We considered {\it data$-$model} approach to generate this profile, where the model represents the best fit straight line at the two different velocity gradient regimes.
It has been suggested that the large-scale velocity oscillations in the pv diagram along the filaments may be related to the kinematics of core formation and to the physical oscillations present in the filaments \citep[e.g.,][see Figure 7 therein]{Liu19}.

In order to compare the velocity oscillations with the column density enhancements in the observed molecular cloud, we have produced the column density profile along a path as shown by a dashed (red) curve in Figure~\ref{fig4}a.
Figure~\ref{fig7}c shows such column density profile, where the zero point in the x-axis corresponds to the eastern end of the path (see Figure~\ref{fig4}a). 
In this exercise, we have used {\it Herschel} multiwavelength images at 160--500 $\mu$m to obtain the column density map (37$''$ resolution) of our target area. The details of the adopted procedure can be found in \citet{Dewangan17N49}.
To make a consistent comparison between velocity and column density plots, we convolved both the data (i.e., $^{13}$CO(1--0) molecular line data and {\it Herschel} column density map) into the same spatial resolution of 38$''$.
In this exercise we used {\texttt CASA}\footnote[1]{https://casa.nrao.edu/} based task {\texttt imsmooth}.
The plots presented in Figure~\ref{fig7} are derived at 38$''$ spatial resolution.
In Figure~\ref{fig7}, we marked vertical dashed lines across all the plots, specifying the positions of local column density peaks.
From the comparison of velocity and column density plots (Figures~\ref{fig7}a and \ref{fig7}c), we note that the velocity peaks are shifted with respect to column density peaks in the direction of G45E. However, this is not obvious in the other parts of the filaments.
The possible outcomes from this analysis are discussed in Section~\ref{sec:edc}.

\subsection{Physical properties of the cloud associated with G45E and G45W}
\label{properties}

\begin{figure*}
\includegraphics[width=0.6\textwidth]{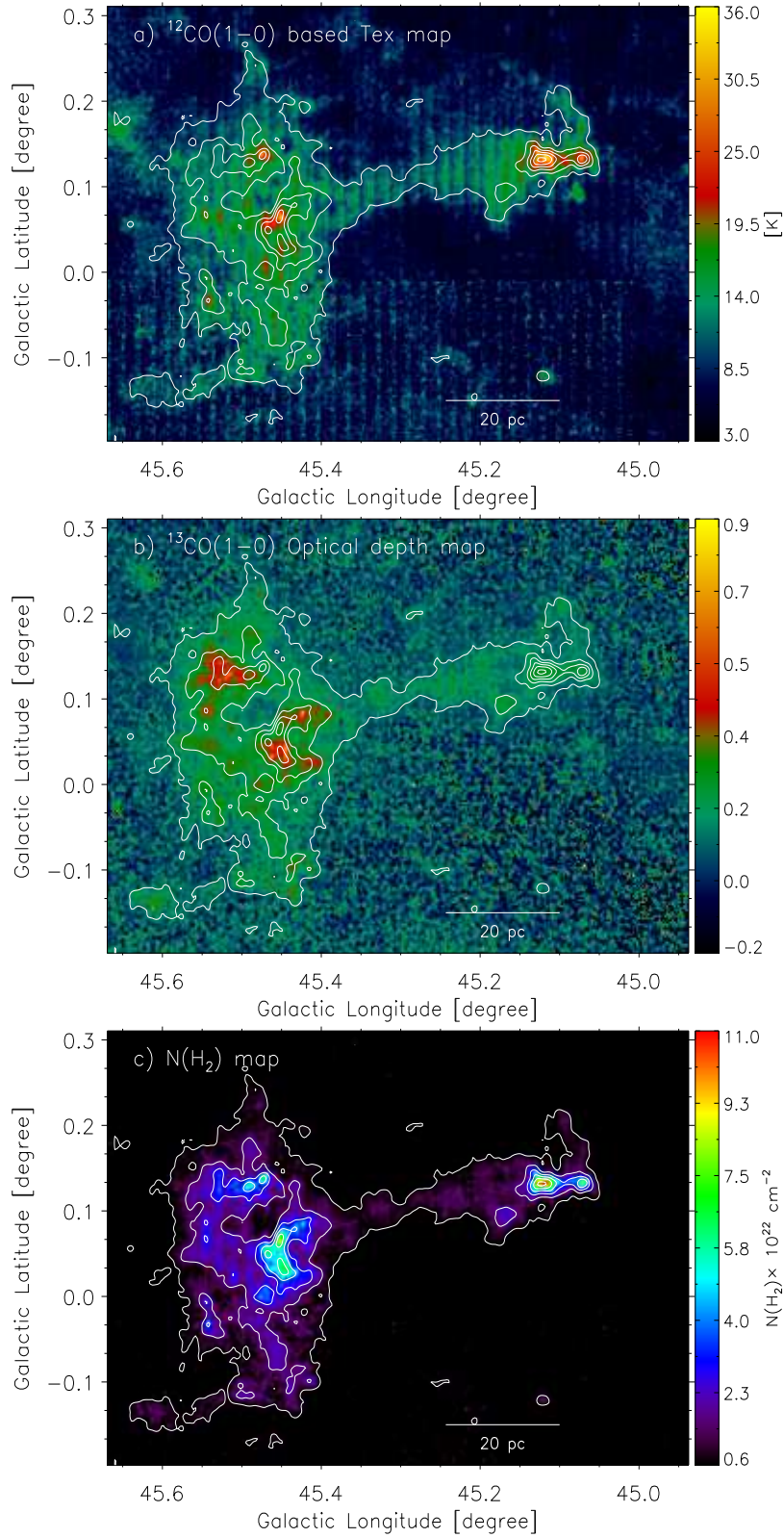}
\caption{
(a) The excitation temperature map derived from $^{12}$CO(1--0) emission in the velocity range of [53, 63] km s$^{-1}$.
(b) $^{13}$CO(1--0) optical depth map.
(c) H$_{2}$ column density map.
In all the panels, the contours represent $^{13}$CO(1--0) integrated emission ranging from 8 to 60 K km s$^{-1}$ in steps of 10.4 K km s$^{-1}$. 
}
\label{fig8}
\end{figure*}

To derive the mass of the filamentary cloud at [53, 63] km s$^{-1}$, we have employed the $^{13}$CO(1--0) line emission.
Considering local thermodynamical equilibrium (LTE), one can estimate the $^{13}$CO(1--0) column density ({\it N}($^{13}$CO)) using the following equation \citep[e.g.,][]{Garden91,Bourke97,Mangum15}

\begin{equation}
N(^{13}{\rm CO})= 2.42\times10^{14}\frac{T_{\rm ex}+0.88}{1-{\rm exp}(-5.29/T_{\rm ex})} \int{\tau_{\rm 13} dv},
\label{13cocolumndensity}
\end{equation}

where $\tau_{\rm 13}$ is the optial depth of $^{13}$CO(1--0), {\it T$_{\rm ex}$} is the mean excitation temperature, and {\it v} is the local standard of rest (LSR) velocity measured in km s$^{-1}$.
Since the $^{12}$CO(1--0) emission is optically thick, one can calculate the T$_{\rm ex}$ using $^{12}$CO(1--0) as follows \citep{Garden91, Xu18}

\begin{equation}
\label{excitationtemp}
T_{\rm ex}= \frac{5.53}{{\rm ln}\left[1+\frac{5.53}{T_{\rm mb}(^{12}{\rm CO})+0.82}\right]},
\end{equation}

where {\it T}$_{\rm mb}$ is the main-beam brightness temperature. 
Figure~\ref{fig8}a presents the excitation temperature map of $^{12}$CO(1--0) emission. The estimated mean excitation temperature is about 8.7 K.
Here, one can assume that the excitation temperatures of $^{12}$CO(1--0) and $^{13}$CO(1--0) are the same for the entire cloud. We can then derive optical depth ($\tau$) using the equation \citep{Garden91, Xu18}

\begin{equation}
\label{tau13co}
\tau(^{13}{\rm CO})=-{\rm ln}\left[1-\frac{T_{\rm mb}(^{13}{\rm CO})}{\frac{5.29}{{\rm exp}(5.29/T_{\rm ex})-1}-0.89}\right]
\end{equation}

The optical depth map of $^{13}$CO(1--0) is shown in Figure~\ref{fig8}b. We found that the optical depth ranges from $\sim$0.1 to 1, suggesting the optically thin nature of the $^{13}$CO emission toward the entire filamentary cloud.
Using Equations~\ref{13cocolumndensity}, \ref{excitationtemp}, and \ref{tau13co}, we have obtained the average value of {\it N}($^{13}$CO) around 2.04$\times$10$^{16}$ cm$^{-2}$. This leads to the mean H$_{\rm 2}$ column density ({\it N}(H$_{\rm 2}$)) of $\sim$1.43$\times$10$^{22}$ cm$^{-2}$ after adopting the column density conversion factor {\it N}(H$_{\rm 2}$)/{\it N}($^{13}$CO) $\simeq$ 7$\times$10$^{5}$ from \citet{Frekring1982}.
In this calculation, we considered the emission inside the filamentary cloud by adopting the minimum {\it N}($^{13}$CO) emission value of 8$\times$10$^{15}$ cm$^{-2}$. The derived {\it N}(H$_{\rm 2}$) map is shown in Figure~\ref{fig8}c.
Here, we also note that the {\it N}(H$_{\rm 2}$) maps derived from the {\it Herschel} continuum images and the molecular line data are consistent with each other. %
We have calculated the mass of the filamentary cloud using the relation of mass and H$_{\rm 2}$ column density as given in Equation~1 of \citet{Bhadari20}. In this calculation, we considered the mean molecular weight of 2.8, and the distance of 8 kpc. 
The total estimated mass of the cloud is $\sim$1.1$\times$10$^{6}$ M$_{\odot}$.
We note that the uncertainties in column density and mass calculations are primarily due to choice of {\it N}(H$_{\rm 2}$)/{\it N}($^{13}$CO) factor, distance estimation, and the observational random errors, which can contribute to 30-50\% uncertainty.
We have also estimated the length of the filamentary cloud to be about 75 pc, which is measured along its long axis (see a dashed (red) curve in Figure~\ref{fig4}a).
The observed line mass of the cloud is then estimated to be $\sim$1.5$\times$10$^{4}$ M$_{\odot}$/pc.
The line mass of the filament depends upon the ``cos {\it i}" factor, where ``{\it i}" is an angle between the sky plane and the major axis of the filament.
In our calculations, we assume {\it i}=0 (i.e., the filament is in the sky plane); hence the derived value represents an upper limit.
However, in addition to the uncertainty in ``{\it i}", the line mass is mainly affected by the uncertainty in distance estimation, which is about 20--30\% (see Section~\ref{sec:intro}).

To assess the gravitational instability of the filamentary cloud, one can compare the observed line mass with the virial line mass. The virial line mass contains the effect of gas turbulence helping the filament to resist the collapse due to gravity. Following the Equation~2 of \citet{Dewangan19S242}, the virial line mass of the entire cloud is found to be $\sim$4.5$\times$10$^{3}$ M$_{\odot}$/pc. In the virial line mass calculation, we considered the expression of non-thermal velocity dispersion ($\sigma_{\rm NT}$) as defined in the Equation~3 of \citet{Bhadari20}, where the observed linewidth ($\Delta$V) of $^{13}$CO(1--0) spectra is used as 6.8 km s$^{-1}$, and the gas kinetic temperature (T$_{\rm K}$) of 18 K. 
The $\Delta$V and T$_{\rm K}$ values are obtained from the Gaussian fitting of averaged $^{13}$CO(1--0) spectra and the dust temperature map for the entire cloud region, respectively.
We have also calculated the line mass parameters of the central region (size $\sim$0$\degr$.29 $\times$ 0$\degr$.056; position angle = 13$\degr$; $l$ = 45$\degr$.29, $b$ = 0$\degr$.109) of the cloud. Concerning to the central region, the observed line mass and the virial line mass are estimated to be $\frac{3.24~\times~10^{4}~{\rm M_{\odot}}}{40~{\rm pc}}$  
$\simeq$800 M$_{\odot}$/pc and $\sim$3.5$\times$10$^{3}$ M$_{\odot}$/pc, respectively. Here, the calcuation uses T$_{\rm K}$ = 18 K and $\Delta$V = 6 km s$^{-1}$.
We note that the uncertainties in these analyses may be relatively high ($\geq$50\%) due to distance uncertainty, an unknown inclination of the filament, and the derived $\Delta$V and T$_{\rm K}$ values. The non-thermal velocity dispersion derived from $\Delta$V could be overestimated due to the optical depth effect and multicomponent structure of the $^{13}$CO(1--0) line.
The implications of these results are discussed in Section~\ref{sec:edc}.

\subsection{Identification and Distribution of YSOs}

The presence of young stellar objects (YSOs) or infrared excess sources in any star-forming region is indicative of ongoing star formation activity.
The origin of infrared emission (in excess) is explained by the radiative reprocessing of starlight by the thick circumstellar material (e.g., envelope, disk) around them. In order to identify the YSOs, we have utilized the photometric data at NIR-MIR wavelengths (1--24 $\mu$m), which are obtained from different infrared surveys (e.g., UKIDSS-GPS, 2MASS, GLIMPSE, and MIPSGAL).
The YSOs towards the target region of our study are identified using the color-color and color-magnitude diagrams.

\begin{figure*}
\includegraphics[width=0.45\textwidth]{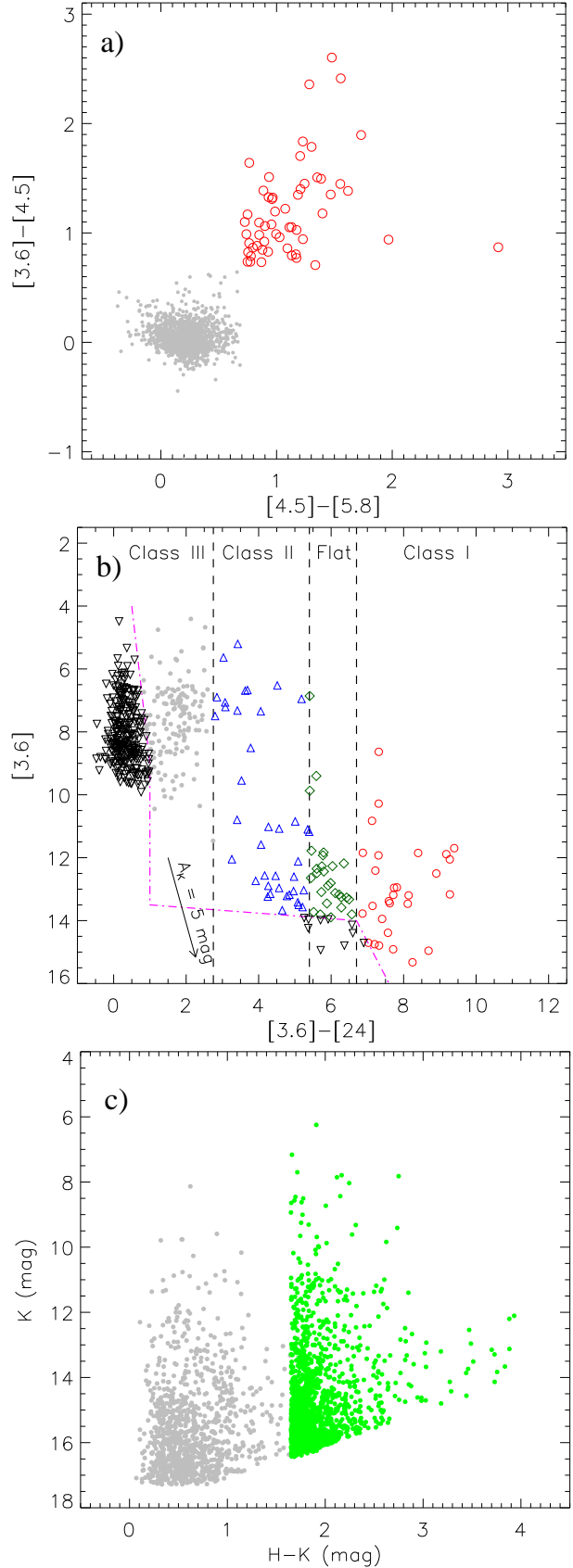}
\caption{(a) Color-color plot ([3.6]--[4.5] vs. [4.5]--[5.8]) of point-like sources in the direction of G045.49+00.04 and G045.14+00.14. (b) Color--magnitude plot ([3.6]--[24] vs. [3.6]) of the sources. The plot enables to identify YSOs belonging to different evolutionary stages (see dashed lines). The boundary of YSOs against contaminated candidates (galaxies and disk-less stars) is shown by dotted-dashed lines (in magenta).
(c) Color--magnitude (H--K vs. K) diagram of point-like sources. The YSO candidate sources are shown by green dots. In panels ``(a)'' and ``(b)'', the Class I sources are marked by red circles. In all the panels, sources with photospheric emission are shown by dots (in gray), which are randomly drawn from a large sample of our survey catalog.
}
\label{fig9}
\end{figure*}

We obtained the photometric magnitudes of point-like objects at 3.6, 4.5, and 5.8 $\mu$m wavelengths from the {\it Spitzer} GLIMPSE-I Spring' 07 highly reliable catalog. In this selection, we considered only objects with a photometric error of less than 0.2 mag in the selected {\it Spitzer} bands.
Figure~\ref{fig9}a shows the color-color plot ([4.5]--[5.8] vs. [3.6]--[4.5]) of point-like objects toward our line of sight.
We followed the infrared color conditions of [4.5]--[5.8] $\geq$ 0.7 mag and [3.6]--[4.5] $\geq$ 0.7 mag from previous works of \citet{Hartmann05} and \citet{Getman07}, and this selection provided us a sample of Class~I YSOs in the extended physical system.
A total of 55 Class~I YSOs are identified using this scheme, which are shown by open red circles in Figure~\ref{fig9}a.

We have also utilized the photometric data at {\it Spitzer} 3.6 and 24 $\mu$m to identify the different classes of YSOs.
Earlier, \citet{Guieu10} used a color–magnitude plot of [3.6]--[24] vs. [3.6] to distinguish different classes of YSOs along with the possible contaminant sources (e.g., galaxies, diskless stars).
Using this scheme, we have identified a total of 87 YSOs (29 Class I, 25 Flat spectrum, and 33 Class II).
The color-magnitude plot of [3.6]--[24] vs. [3.6] is shown in Figure~\ref{fig9}b, where the Class~I, Flat spectrum, and Class~II sources are indicated by red circles, green diamonds, and blue triangles, respectively. 

\begin{figure*}
\includegraphics[width=\textwidth]{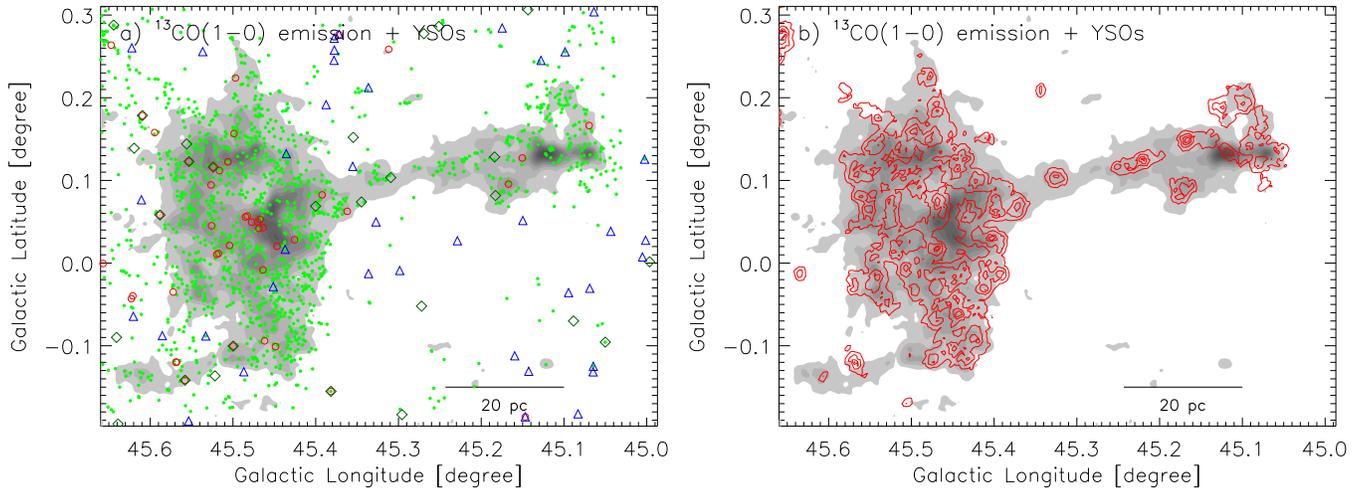}
\caption{(a) The spatial distribution of YSOs (from Figure~\ref{fig9}) overlaid on the $^{13}$CO(1--0) integrated intensity map (see Figure~\ref{fig4}a). The symbols are same as discussed in Figure~\ref{fig9}.
(b) Overlay of the YSO surface density contours (in red) on the $^{13}$CO(1--0) emission map.
The contour values are 0.6, 1.1, 2, 3 YSOs pc$^{-2}$.
}
\label{fig10}
\end{figure*}

Additional color excess sources are also obtained from the color-magnitude diagram of $H-K$ vs. $K$ (see Figure~\ref{fig9}c). 
A large number of 1486 color excess sources (marked by green dots) are identified following the color cutoff condition of $H-K$ $>$ 1.65. This condition is decided based upon the upper limit of $H-K$ detection of main sequence stars in the nearby control field. 
Altogether, after taking account of the common sources identified from different schemes, a total of 1628 YSO candidate sources are 
identified toward our selected target area. The positions of all the YSOs are presented in Figure~\ref{fig10}a.

We have also performed the surface density analysis of all the selected YSO candidates using the nearest neighbour (NN) method \citep[e.g.,][]{Gutermuth2009}.
To generate the surface density map, we considered a grid size of 15$''$ and the 6 NN YSOs at the distance of 8 kpc \citep[see formalism in][]{Casertano1985}.
Figure~\ref{fig10}b presents the overlay of surface density contours (in red) on the $^{13}$CO(1--0) integrated intensity map. 
The surface density distribution is primarily concentrated at the positions of G45E and G45W.

\subsection{Structure of the dense gas and hub-filament systems}
\label{hfs}

\begin{figure*}
\includegraphics[width=0.9\textwidth]{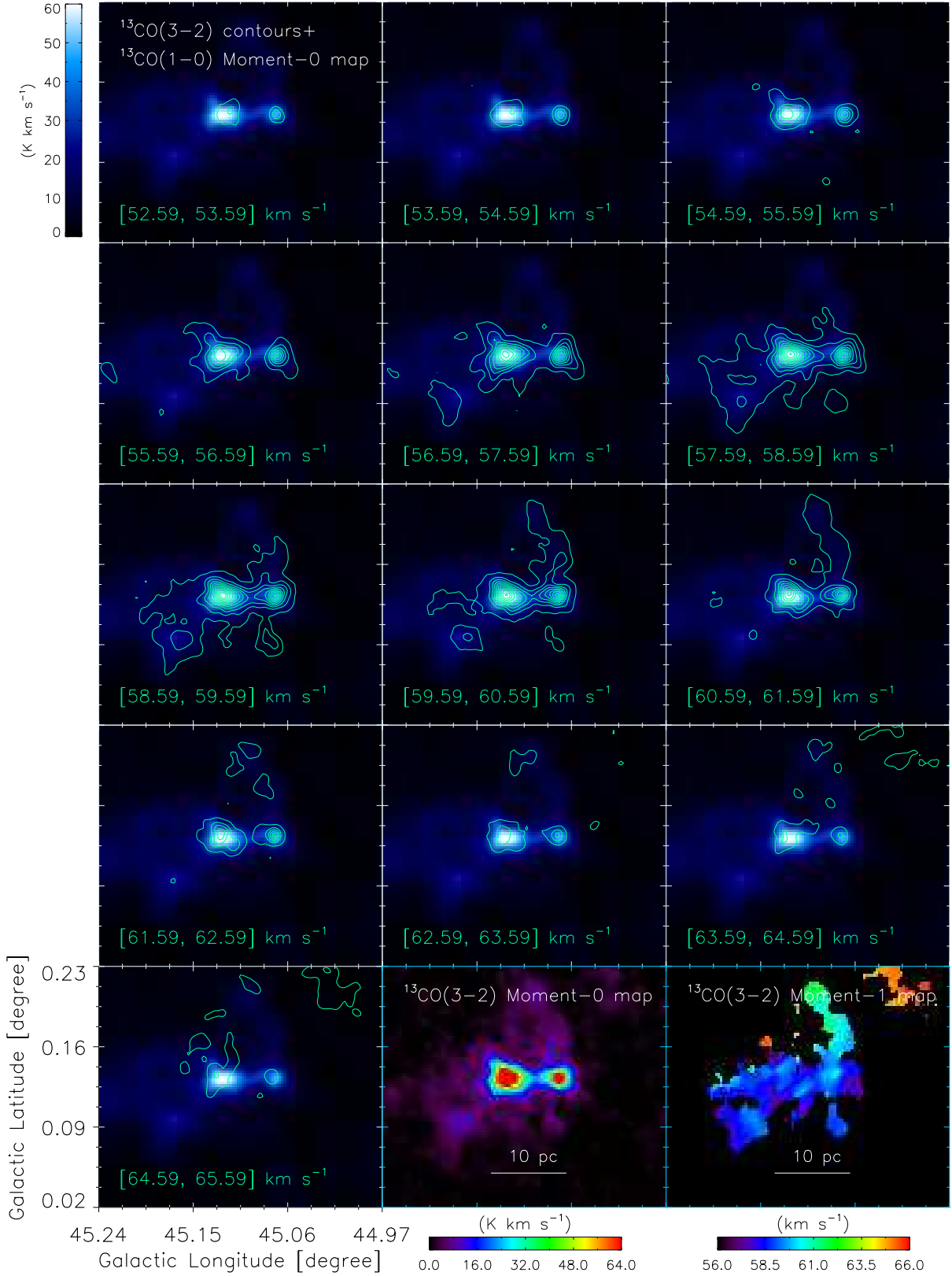}
\caption{Velocity channel maps of $^{13}$CO(3--2) emission (green contours) in the direction of G045.14+00.14 (i.e., G45W) region. 
The contour levels range from 1.2 to 26 K km s$^{-1}$ in steps of 2.48 K km s$^{-1}$.
The background image is $^{13}$CO(1--0) moment-0 map obtained in the velocity range of [53, 63] km s$^{-1}$, and the color bar is shown at the top left corner.
The last two panels display the moment-0 and moment-1 maps of $^{13}$CO(3--2) emission with their respective color bars.
The velocities are labeled in each panel.
}
\label{fig11}
\end{figure*}

We have used $^{13}$CO(3--2) line data to trace the morphology and kinematics of dense gas. The velocity channel maps of $^{13}$CO(3--2) emission in the direction of G45W are shown in Figure~\ref{fig11}.
We have overlaid the velocity averaged emission (at each 1 km interval; see contours) on the $^{13}$CO(1--0) moment-0 map.
This exercise allows us to identify the dense gas structures, which are highly elongated/filamentary as seen in the velocity range from 55.59 km s$^{-1}$ to 61.59 km s$^{-1}$. The elongated gas structures are extended to the locations of both the IRAS sources (i.e., IRAS 19110+1045 and IRAS 19111+1048). The last two panels of the channel maps present the moment-0 and moment-1 maps of the $^{13}$CO(3--2) emission, which are obtained for the velocity range as presented in the channel maps (i.e., [52.59, 64.59] km s$^{-1}$). 
From the moment maps, the presence of elongated structures and the velocity gradient along them are evident.

Based on the visual inspection of Figure~\ref{fig2}a, we can notice the possible presence of filamentary structures (subfilaments) in our target site.
To further explore these structures, we have used {\it getsf}, a newly developed tool for extraction of sources and filaments in astronomical images \citep{Men2021}. The {\it getsf} spatially decomposes the image into its structural components and separates them from each other and from their backgrounds. 
After flattening the residual noise in the resulting images, it decomposes the flattened images and identifies the positions of sources and skeletons of filaments.
To identify structures using {\it getsf}, it requires a single user-defined input i.e., the maximum size of the structure to be extracted (e.g., filaments in our case).

\begin{figure*}
\includegraphics[width=\textwidth]{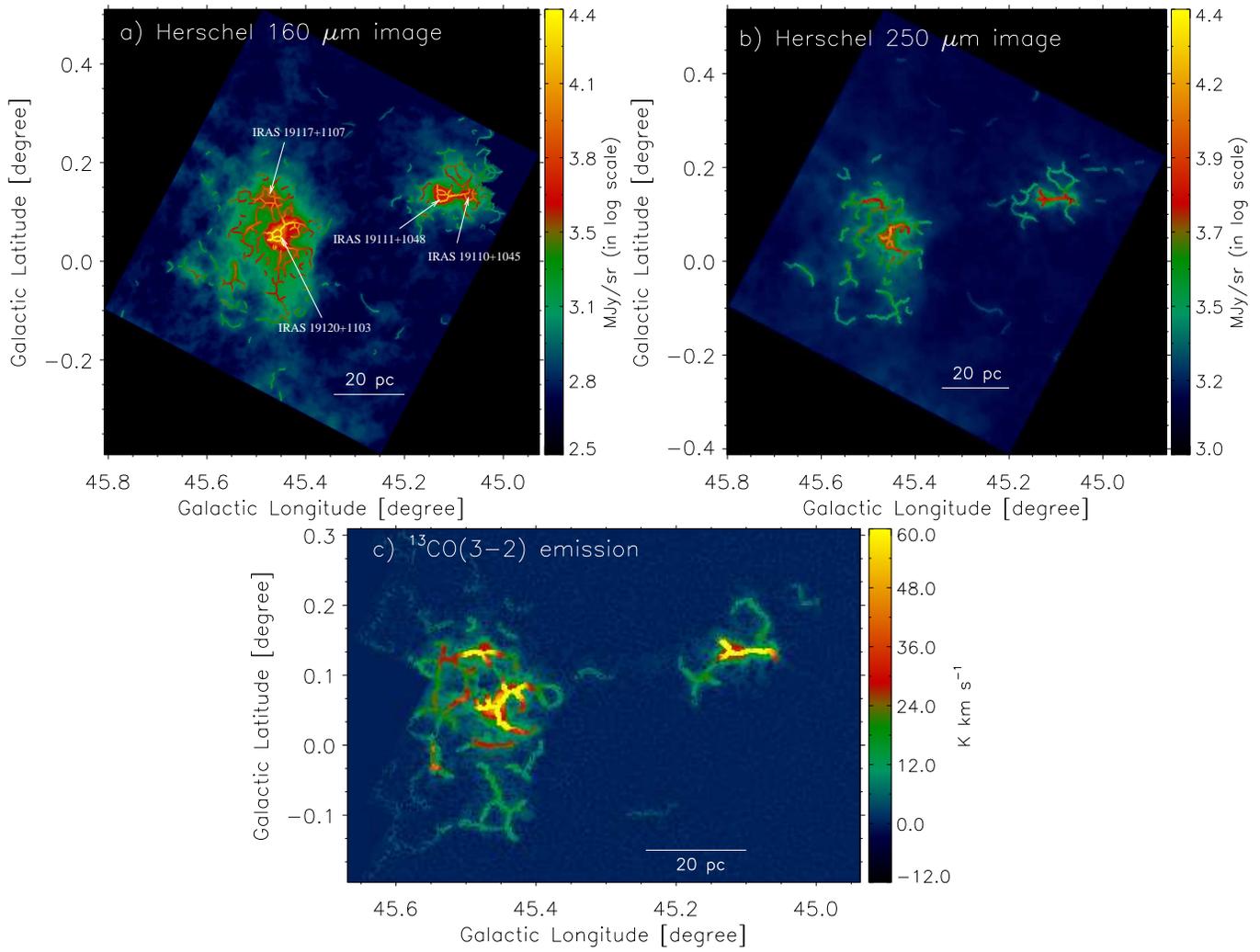}
\caption{Overlay of the {\it getsf} identified filament skeletons on the {\it Herschel} continuum images at (a) 160 $\mu$m and (b) 250 $\mu$m, and on the (c) CHIMPS $^{13}$CO(3--2) moment-0 map (see text for more details). The positions of the IRAS sources are marked by arrows in panel (a).
}
\label{fig12}
\end{figure*}

Figure~\ref{fig12} presents the {\it getsf} identified filament skeletons (on global scale) in our target site.
We have used the {\it Herschel} continuum images at 160 $\mu$m and 250 $\mu$m to identify the subfilaments. Figure~\ref{fig12}a shows the {\it Herschel} 160 $\mu$m filament skeletons identified on the image scales of 12$''$-543$''$. 
Figure~\ref{fig12}b displays the {\it Herschel} 250 $\mu$m filament skeletons identified on the scales of 18$''$-576$''$. 
In order to trace the genuine subfilaments in a star-forming site, one should also look for filaments in molecular line emission apart from the continuum maps. The {\it getsf} procedure is also applied to the CHIMPS $^{13}$CO(3--2) moment-0 map (V$_{lsr}$ $\sim$53--66 km s$^{-1}$, $\theta_{\rm spatial}$=15$''$). In Figure~\ref{fig12}c, the {\it getsf} identified filaments on the scales of 15$''$-679$''$ are highlighted on the CHIMPS $^{13}$CO(3--2) moment-0 map. 
In the whole exercise, the input parameter of maximum width (FWHM, in arcsec) of filaments to be extracted was used as 260$''$, 420$''$, and 400$''$ for the {\it Herschel} 160 $\mu$m, the {\it Herschel} 250 $\mu$m, and the CHIMPS $^{13}$CO(3--2) emission maps, respectively.

From the comparison of maps displayed in Figure~\ref{fig12}, we notice that the filamentary structures in the direction of dense regions (i.e., high intensity regions) are consistent with the structures seen in the continuum and molecular emission maps, thus revealing the potential subfilaments in our target sites. 
It is interesting to note that the structures of subfilaments in sites G45E and G45W are the reminiscences of HFSs (see Section~\ref{sec:intro}). 
The hub-filament configurations are clearly evident in both the complexes (i.e., toward G45E and G45W), which are primarily associated with the previously known massive star-forming sites (i.e., IRAS sources; see Figure~\ref{fig12}a).
These outcomes are further discussed in Section~\ref{sec:hfs}.

\section{Discussion}
\label{sec:discussion}

\subsection{End dominated collapse in a giant molecular filament GMF G45.3+0.1}
\label{sec:edc}

In Section~\ref{kinematics_g45}, we have investigated that the sites G45E and G45W are part of a filamentary cloud (see also Section~\ref{sec:intro}), and are located at the ends of it.
Based on the physical length of 75 pc and the derived mass of $\sim$1.1$\times$10$^{6}$ M$_{\odot}$ (see Section~\ref{properties}), the cloud can be considered as a giant molecular cloud \citep[GMC;][]{Solomon1987}. Hence, considering the position in the Galactic system, we named this filamentary cloud as a giant molecular filament (GMF) G45.3+0.1. 
However, previously the eastern part of GMF G45.3+0.1 (i.e., G45E) was itself considered to be a large filament and cataloged as ``F39" by \citet[][see Figure~6 therein]{Wang2016}. \citet{Zhang19} further included the western part of GMF G45.3+0.1 (i.e., G45W) and cataloged the entire filament as ``GMF 37".
The clear definition of a GMF is not available in the literature. However, according to \citet{Zhang19}, one must see a clear elongation in the cloud over some column-density regime, and the cloud should show coherence in the position-position-velocity (PPV) space. The GMFs have lengths of the order of $\sim$100 pc and masses of $\sim$10$^{3}$-10$^{6}$ M$_{\odot}$ \citep[e.g.,][]{Li2013,Ragan2014,Su2015,Abreu2016,Zhang19}.
Adopting a typical cloud width range of $\sim$8--10 pc, which can vary along the major axis of the cloud because of its evolution in terms of structure, shaped by the feedback from embedded stars
\citep[e.g.,][]{McKee2007}, the GMF G45.3+0.1, always have a high aspect ratio (A$\sim$7--10; i.e., a ratio of filament's length to diameter). 
Such high aspect ratio (A$>$5) filaments are considered to collapse at the edges due to the gravitational instability induced by the high gas acceleration \citep{Bastien83,Pon12,Clarke15,Hoemann2022arXiv}.

Till now, very few observational examples of filaments showing end-dominated collapse are available in the literature, which include NGC 6334 \citep{Zernickel2013}, IRDC~18223 \citep{Beuther2015}, Musca cloud \citep{Kainulainen2016}, S242 \citep{Dewangan2017,Dewangan19S242}, G341.244$-$00.265 \citep{Yu2019}, and Monoceros R1 \citep{Bhadari20}.
The key features of such physical process include at least one of the following possibilities i.e., the existence of high column density clumps, star clusters, and H{\sc ii} regions at one or both the filament ends. Such filaments should ideally be observed as isolated clouds in 3D or position-position-position (PPP) space.
However, these filaments are generally identified as coherent velocity clouds in the PPV space.
Theoretically, there is no restriction on the filament's parameter (i.e., length or radius) for it to collapse by the end-dominated process; instead, the global collapse of filament (specifically, the collapse timescale) is dependent on the aspect ratio \citep[e.g.,][]{Toala2012,Pon12,Clarke15}. 
In Section~\ref{properties}, we found that the observed line mass of GMF G45.3+0.1 is larger than the virial line mass, suggesting that it is gravitationally unstable and prone to collapse. However, the central region of GMF G45.3+0.1, where the star formation activity is not dominated, shows the opposite nature. Previously, \citet{Dewangan19S242} also found that the central region of S242 filament shows higher virial mass compared to the observed line mass. This suggests that the cloud is self-gravitating and gravitational unstable at the edges compared to the central areas.

Based on our observational outcomes, we argue that the global collapse of GMF G45.3+0.1 is predominately end-dominated in nature.
This argument is supported by the fact that the sites G45E and G45W are observed at the ends of GMF G45.3+0.1, where a large number of YSO clusterings and the presence of H{\sc ii} regions are also observed. 
The major column density peaks are also seen at the ends of GMF G45.3+0.1 (Figure~\ref{fig7}c).
We also note that the complex G45E is more extended than the G45W. It is probably possible because of the rapid structural evolution of G45E from the feedback of embedded stars.

The fragmentation of GMF G45.3+0.1 appears to proceed in a hierarchy (i.e., into subfilaments) as seen in the dense gas site G45W.
In the direction of complex G45W, the ATLASGAL map at 870 $\mu$m shows the presence of a filamentary structure (length $\sim$12 pc; see Figure~\ref{fig2}c) connecting the two massive star-forming sites, IRAS 19110+1045 and IRAS 19111+1048.
This subfilament is found to be denser than the parent filament, as it is traced in the relatively dense gas tracers of $^{13}$CO/C$^{18}$O(3--2) emission (see Figures~\ref{fig4}b,~\ref{fig11}), thus indicating the hierarchical fragmentation of GMF G45.3+0.1. 
Some previous observational \citep[e.g.,][]{Hacar2013,Tafalla2015,Hacar2018} and theoretical \citep[e.g.,][]{Smith2016,Clarke2017,Clarke2020} works also favor the hierarchical fragmentation of filaments. These results significantly deviate from previous models of filament fragmentation which favour the formation of quasi-periodically spaced cores along filaments \citep[e.g.,][]{Inutsuka1992,Inutsuka1997,Fischera2012}. 
\citet{Clarke2020} argue that the hierarchical fragmentation of filaments diminishes the signature of characteristic fragmentation length-scale (related to the quasi-periodically spaced cores), and the cores are formed by the fragmentation of subfilaments rather than the main filament itself.
We observed oscillations in the mean velocity profile along the long axis of GMF G45.3+0.1 (see Figures~\ref{fig6}c and \ref{fig7}), which hints the signature of core formation \citep[e.g.,][]{Beuther2015,Kainulainen2016} together with the possibility of large scale physical oscillations in the filament \citep{Liu19}.
However, based on the comparison of velocity and column density plots (Figure~\ref{fig7}), we observed that the column density and velocity peaks toward G45E are shifted with respect to each other. This could be an indication of the presence of gravitationally bound cores \citep[e.g.,][see Figure 12 therein]{Hacar2011}. Some of the other column density peaks along filament do not significantly correlate to the velocity oscillations, suggesting that the cores are not gravitationally bound.

We have observed linear gradients in the velocity profile along the long axis of GMF G45.3+0.1.
At the eastern end of the filament (toward G45E), the linear gradient is negative and twice in magnitude to that of the positive gradient toward the western end.
The origin of these adverse gradients at the filament's edges can probably be explained by the presence of sites G45E and G45W itself, where the feedback from massive stars in these sites appears to shape the cloud in terms of structure and kinematics.

\subsection{Hub-filament systems and star formation scenario}
\label{sec:hfs}

The current understanding of star formation argues that the complex filaments form initially in the molecular clouds \citep[e.g.,][and references therein]{Chen2020MNRAS.494.3675C,Abe21}, which then undergo the fragmentation processes to form dense clumps/cores and ultimately the stellar clusters \citep[e.g.,][]{Hacar2013}. 
However, despite taking birth in the clustered environment, the formation theories of massive stars ($\geq$8~M$_\odot$) are still an enigmatic topic of research \citep[e.g.,][and references therein]{Motte2018}.
In this context, the necessity of the mass reservoir becomes vital for the formation of massive stars (i.e., massive star formation; MSF), which can be fulfilled in the physical environment of filaments.

In Section~\ref{hfs}, we found that the star-forming complexes G45E and G45W appear to have the morphology of parsec scale HFSs.
These HFSs are primarily investigated in the direction of massive star-forming regions (i.e., IRAS sources; see Figures~\ref{fig2} and \ref{fig12}), and have the filament's length of the order of $\sim$5--10 pc in the vicinity of IRAS sites.
Overall, these filaments show converging networks toward the hubs (or local intensity peaks), where they either direct to the hubs or merge with the other longer spines approaching to hubs. 
The onset of the edge-collapse process, together with the presence of HFSs in GMF G45.3+0.1, is quite intriguing, hence opens up a new possibility of MSF, including the formation of HFSs \citep[e.g.,][and references therein]{Motte2018,Wang2020}.

The EDC process can well explain the formation of star-forming sites G45E and G45W. However, the presence of HFSs in these sites itself hints the onset of global nonisotropic collapse \citep[GNIC;][]{Tige2017, Motte2018,Morales2019}.
GNIC scenario supports the mass accretion onto the dense core from the large scale structures.
In this paradigm, a density-enhanced hub hosts massive dense cores (MDCs; size$\sim$0.1 pc), which in their starless phase harbor low-mass prestellar cores.
Later, MDCs become protostellar, hosting only low-mass stellar embryos, which grow into high-mass protostars from gravitationally-driven
inflows \citep[see more details in][]{Motte2018}.
Observational studies of HFSs report that the longitudinal flow rate along filaments is around 10$^{-4}$--10$^{-3}$~M$_\odot$ year$^{-1}$ \citep[e.g.,][]{Kirk2013,Chen2019,Morales2019} which is a close value to the sufficient inflow rate for MSF \citep[e.g.,][and references therein]{Haemmerle2016}. 
Hence, it is now believed that all massive stars favour forming in the density-enhanced hubs of HFSs \citep[e.g.,][and references therein]{Motte2018,Anderson2021}. 
It has also been thought that the onset of collision process \citep[see the review article by][and references therein]{fukui21} can produce hubs with large networks of filaments. 
The filamentary structures, reminiscent of the HFSs are quite evident in the smoothed particle hydrodynamics simulations of two colliding clouds \citep{Balfour2015}.
In this case, the shock-compressed layer is found to fragment into filaments. Furthermore, the pattern of filaments appears to have the morphology of HFSs, spider's web, and spokes system \citep{Balfour2015}. 


Recently, \citet{Clarke2020} found that the filaments are more prone to fragment into subfilaments, and these subfilaments are responsible for the formation of hubs via their merging event. Also, the cores which are formed at the filament's edges are more massive than the interior ones \citep{Clarke2020}.
Thus, one can argue that the HFSs seen in a giant molecular filament are possibly formed by the combined effect of colliding subfilaments and the fragmentation of shock-compressed layers in the collision sites.
Comparing all these theoretical results with our observational outcomes of the study of GMF G45.3+0.1, we argue that the star formation activity in G45E and G45W, primarily proceed through the onset of EDC and HFSs.
We notice that the HFSs are preferentially formed at the edges of GMF G45.3+0.1, where massive end clumps are also expected from the EDC process.
All our observations are constrained by the limited resolution of the existing data, which for a large source distance of our target site (8 kpc for GMF G45.3+0.1), make it further intricate.

Thus, our observational outcomes reveal GMF G45.3+0.1 as an observational sample of filament, identified so far in the interstellar medium, where the edge collapse and the hub-filament configurations are simultaneously seen.
Our observational results are more or less consistent with the outcomes from the numerical fragmentation study on filaments by \citet{Clarke2020}.
Such a study also helps us to understand the formation of HFSs and MSF.

\section{Summary and Conclusions}
\label{sec:conc}
This paper is aimed to get new insights into the physical processes operating in the sites G45E and G45W, which are part of a large cloud complex in the Aquila constellation ($l$=45$^{o}$, $b$=0$^{o}$).
We have performed the analysis of multiwavelength and multiscale data-sets toward an area (0$\degr$.67$\times$0$\degr$.50) enclosing our target sites.
The major outcomes from our observational study can be summarized as follows: 

\begin{enumerate}

\item The analysis of $^{13}$CO(1--0) line data reveals the presence of a filamentary cloud (length $\sim$75 pc) in the velocity range of $\sim$53--63 km s$^{-1}$ having the mean velocity and standard deviation of 58.30$\pm$0.03 km s$^{-1}$ and 2.34$\pm$0.03 km s$^{-1}$, respectively.

\item Using $^{13}$CO(1--0) emission maps, the cloud mass is derived to be $\sim$1.1$\times$10$^{6}$ M$_{\odot}$, which further results in the line mass of $\sim$1.5$\times$10$^{4}$ M$_{\odot}$/pc. 
Our study thus uncovers a new giant molecular filament, which we named GMF G45.3+0.1.

\item The sites G45E and G45W are observed at the opposite ends of GMF G45.3+0.1. The clustering of YSOs is primarily focused at both ends of the filament, where the presence of H{\sc ii} regions and previously known massive star-forming sites 
are also observed.

\item The pv diagram along the major axis of GMF G45.3+0.1 shows the velocity oscillations, which could be originated by the combined effect of fragment/core formation and the physical oscillation in the filament.
The velocity gradient at both ends of the filament is found to be opposite in sign, i.e., $-$0.064 km s$^{-1}$ pc$^{-1}$ toward G45E and $+$0.032 km s$^{-1}$ pc$^{-1}$ in the direction of G45W.

\item The sites IRAS 19110+1045 and IRAS 19111+1048, located in G45W, are primarily observed to be a part of $\sim$12 pc long dense subfilament. This presents strong evidence of hierarchical fragmentation of GMF G45.3+0.1.

\item The analysis of the dust continuum and molecular line data suggests that both the complexes (G45E and G45W) are associated with the HFSs, where several parsec scale filaments ($\sim$5--10 pc) appear to converge toward dense regions (i.e., hubs). These hubs are identified in the direction of massive star-forming sites (i.e., IRAS 19120+1103, IRAS 19117+1107, IRAS 19110+1045, and IRAS 19111+1048).

\end{enumerate}

Overall, our observational results suggest that the star-forming complexes in the GMF G45.3+0.1 are formed through the edge collapse scenario and the mass accumulation
is evident in each hub through the gas flows along filaments.

\section*{Acknowledgements}
We thank the referee for valuable comments and suggestions that helped improve the paper's presentation.
The research work at Physical Research Laboratory is funded by the Department of Space, Government of India. 
DKO acknowledges the support of the Department of Atomic Energy, Government of India, under Project Identification No. RTI 4002.
LEP acknowledges the support of the IAP State Program 0030-2021-0005.
NKB thanks Peter Zemlyanukha for helping in the cloud mass calculation using molecular line data.
NKB extends heartfelt gratitude to Alexander Men'shchikov for the discussion on the {\it getsf} tool and its outputs.
This work is based on data obtained as part of the UKIRT Infrared Deep Sky Survey. 
This publication makes use of data products from the Two Micron All Sky Survey, which is a joint project of the University of Massachusetts and the Infrared Processing and Analysis Center/California Institute of Technology, funded by NASA and NSF.
This work is based [in part] on observations made with the {\it Spitzer} Space Telescope, which is operated by the Jet Propulsion Laboratory, California Institute of Technology under a contract with NASA. This publication makes use of data from FUGIN, FOREST Unbiased Galactic plane Imaging survey with the Nobeyama 45-m telescope, a legacy project in the Nobeyama 45-m radio telescope.


\bibliography{reference}{}

\begin{thebibliography}{}
\expandafter\ifx\csname natexlab\endcsname\relax\def\natexlab#1{#1}\fi
\providecommand{\url}[1]{\href{#1}{#1}}
\providecommand{\dodoi}[1]{doi:~\href{http://doi.org/#1}{\nolinkurl{#1}}}
\providecommand{\doeprint}[1]{\href{http://ascl.net/#1}{\nolinkurl{http://ascl.net/#1}}}
\providecommand{\doarXiv}[1]{\href{https://arxiv.org/abs/#1}{\nolinkurl{https://arxiv.org/abs/#1}}}

\bibitem[{{Abe} {et~al.}(2021){Abe}, {Inoue}, {Inutsuka}, \&
  {Matsumoto}}]{Abe21}
{Abe}, D., {Inoue}, T., {Inutsuka}, S.-i., \& {Matsumoto}, T. 2021, \apj, 916,
  83, \dodoi{10.3847/1538-4357/ac07a1}

\bibitem[{{Abreu-Vicente} {et~al.}(2016){Abreu-Vicente}, {Ragan},
  {Kainulainen}, {Henning}, {Beuther}, \& {Johnston}}]{Abreu2016}
{Abreu-Vicente}, J., {Ragan}, S., {Kainulainen}, J., {et~al.} 2016, \aap, 590,
  A131, \dodoi{10.1051/0004-6361/201527674}

\bibitem[{{Anderson} {et~al.}(2021){Anderson}, {Peretto}, {Ragan}, {Rigby},
  {Avison}, {Duarte-Cabral}, {Fuller}, {Shirley}, {Traficante}, \&
  {Williams}}]{Anderson2021}
{Anderson}, M., {Peretto}, N., {Ragan}, S.~E., {et~al.} 2021, \mnras,
  \dodoi{10.1093/mnras/stab2674}

\bibitem[{{Andr{\'e}} {et~al.}(2010){Andr{\'e}}, {Men'shchikov}, {Bontemps},
  {K{\"o}nyves}, {Motte}, {Schneider}, {Didelon}, {Minier}, {Saraceno},
  {Ward-Thompson}, {di Francesco}, {White}, {Molinari}, {Testi}, {Abergel},
  {Griffin}, {Henning}, {Royer}, {Mer{\'\i}n}, {Vavrek}, {Attard},
  {Arzoumanian}, {Wilson}, {Ade}, {Aussel}, {Baluteau}, {Benedettini},
  {Bernard}, {Blommaert}, {Cambr{\'e}sy}, {Cox}, {di Giorgio}, {Hargrave},
  {Hennemann}, {Huang}, {Kirk}, {Krause}, {Launhardt}, {Leeks}, {Le Pennec},
  {Li}, {Martin}, {Maury}, {Olofsson}, {Omont}, {Peretto}, {Pezzuto}, {Prusti},
  {Roussel}, {Russeil}, {Sauvage}, {Sibthorpe}, {Sicilia-Aguilar}, {Spinoglio},
  {Waelkens}, {Woodcraft}, \& {Zavagno}}]{Andre10}
{Andr{\'e}}, P., {Men'shchikov}, A., {Bontemps}, S., {et~al.} 2010, \aap, 518,
  L102, \dodoi{10.1051/0004-6361/201014666}

\bibitem[{{Balfour} {et~al.}(2015){Balfour}, {Whitworth}, {Hubber}, \&
  {Jaffa}}]{Balfour2015}
{Balfour}, S.~K., {Whitworth}, A.~P., {Hubber}, D.~A., \& {Jaffa}, S.~E. 2015,
  \mnras, 453, 2471, \dodoi{10.1093/mnras/stv1772}

\bibitem[{{Bastien}(1983)}]{Bastien83}
{Bastien}, P. 1983, \aap, 119, 109

\bibitem[{{Benjamin} {et~al.}(2003){Benjamin}, {Churchwell}, {Babler}, {Bania},
  {Clemens}, {Cohen}, {Dickey}, {Indebetouw}, {Jackson}, {Kobulnicky},
  {Lazarian}, {Marston}, {Mathis}, {Meade}, {Seager}, {Stolovy}, {Watson},
  {Whitney}, {Wolff}, \& {Wolfire}}]{Benjamin03}
{Benjamin}, R.~A., {Churchwell}, E., {Babler}, B.~L., {et~al.} 2003, \pasp,
  115, 953, \dodoi{10.1086/376696}

\bibitem[{{Beuther} {et~al.}(2015){Beuther}, {Ragan}, {Johnston}, {Henning},
  {Hacar}, \& {Kainulainen}}]{Beuther2015}
{Beuther}, H., {Ragan}, S.~E., {Johnston}, K., {et~al.} 2015, \aap, 584, A67,
  \dodoi{10.1051/0004-6361/201527108}

\bibitem[{{Bhadari} {et~al.}(2020){Bhadari}, {Dewangan}, {Pirogov}, \&
  {Ojha}}]{Bhadari20}
{Bhadari}, N.~K., {Dewangan}, L.~K., {Pirogov}, L.~E., \& {Ojha}, D.~K. 2020,
  \apj, 899, 167, \dodoi{10.3847/1538-4357/aba2c6}

\bibitem[{{Blum} \& {McGregor}(2008)}]{Blum08}
{Blum}, R.~D., \& {McGregor}, P.~J. 2008, \aj, 135, 1708,
  \dodoi{10.1088/0004-6256/135/5/1708}

\bibitem[{{Bourke} {et~al.}(1997){Bourke}, {Garay}, {Lehtinen},
  {K{\"o}hnenkamp}, {Launhardt}, {Nyman}, {May}, {Robinson}, \&
  {Hyland}}]{Bourke97}
{Bourke}, T.~L., {Garay}, G., {Lehtinen}, K.~K., {et~al.} 1997, \apj, 476, 781,
  \dodoi{10.1086/303642}

\bibitem[{{Carey} {et~al.}(2005){Carey}, {Noriega-Crespo}, {Price}, {Padgett},
  {Kraemer}, {Indebetouw}, {Mizuno}, {Ali}, {Berriman}, {Boulanger}, {Cutri},
  {Ingalls}, {Kuchar}, {Latter}, {Marleau}, {Miville-Deschenes}, {Molinari},
  {Rebull}, \& {Testi}}]{Carey05}
{Carey}, S.~J., {Noriega-Crespo}, A., {Price}, S.~D., {et~al.} 2005, in
  American Astronomical Society Meeting Abstracts, Vol. 207, American
  Astronomical Society Meeting Abstracts, 63.33

\bibitem[{{Casertano} \& {Hut}(1985)}]{Casertano1985}
{Casertano}, S., \& {Hut}, P. 1985, \apj, 298, 80, \dodoi{10.1086/163589}

\bibitem[{{Chen} {et~al.}(2020{\natexlab{a}}){Chen}, {Mundy}, {Ostriker},
  {Storm}, \& {Dhabal}}]{Chen2020MNRAS.494.3675C}
{Chen}, C.-Y., {Mundy}, L.~G., {Ostriker}, E.~C., {Storm}, S., \& {Dhabal}, A.
  2020{\natexlab{a}}, \mnras, 494, 3675, \dodoi{10.1093/mnras/staa960}

\bibitem[{{Chen} {et~al.}(2019){Chen}, {Zhang}, {Wright}, {Busquet}, {Lin},
  {Liu}, {Olguin}, {Sanhueza}, {Nakamura}, {Palau}, {Ohashi}, {Tatematsu}, \&
  {Liao}}]{Chen2019}
{Chen}, H.-R.~V., {Zhang}, Q., {Wright}, M.~C.~H., {et~al.} 2019, \apj, 875,
  24, \dodoi{10.3847/1538-4357/ab0f3e}

\bibitem[{{Chen} {et~al.}(2020{\natexlab{b}}){Chen}, {Di Francesco},
  {Rosolowsky}, {Keown}, {Pineda}, {Friesen}, {Caselli}, {Chen}, {Matzner},
  {Offner}, {Punanova}, {Redaelli}, {Scibelli}, \& {Shirley}}]{Chen2020ApJ}
{Chen}, M. C.-Y., {Di Francesco}, J., {Rosolowsky}, E., {et~al.}
  2020{\natexlab{b}}, \apj, 891, 84, \dodoi{10.3847/1538-4357/ab7378}

\bibitem[{{Clarke} \& {Whitworth}(2015)}]{Clarke15}
{Clarke}, S.~D., \& {Whitworth}, A.~P. 2015, \mnras, 449, 1819,
  \dodoi{10.1093/mnras/stv393}

\bibitem[{{Clarke} {et~al.}(2017){Clarke}, {Whitworth}, {Duarte-Cabral}, \&
  {Hubber}}]{Clarke2017}
{Clarke}, S.~D., {Whitworth}, A.~P., {Duarte-Cabral}, A., \& {Hubber}, D.~A.
  2017, \mnras, 468, 2489, \dodoi{10.1093/mnras/stx637}

\bibitem[{{Clarke} {et~al.}(2020){Clarke}, {Williams}, \& {Walch}}]{Clarke2020}
{Clarke}, S.~D., {Williams}, G.~M., \& {Walch}, S. 2020, \mnras, 497, 4390,
  \dodoi{10.1093/mnras/staa2298}

\bibitem[{{Dewangan}(2021)}]{Dewangan2021MNRAS}
{Dewangan}, L.~K. 2021, \mnras, 504, 1152, \dodoi{10.1093/mnras/stab1008}

\bibitem[{{Dewangan} {et~al.}(2017{\natexlab{a}}){Dewangan}, {Baug}, {Ojha},
  {Janardhan}, {Devaraj}, \& {Luna}}]{Dewangan2017}
{Dewangan}, L.~K., {Baug}, T., {Ojha}, D.~K., {et~al.} 2017{\natexlab{a}},
  \apj, 845, 34, \dodoi{10.3847/1538-4357/aa7da2}

\bibitem[{{Dewangan} {et~al.}(2017{\natexlab{b}}){Dewangan}, {Ojha}, \&
  {Baug}}]{Dewangan2017b}
{Dewangan}, L.~K., {Ojha}, D.~K., \& {Baug}, T. 2017{\natexlab{b}}, \apj, 844,
  15, \dodoi{10.3847/1538-4357/aa79a5}

\bibitem[{{Dewangan} {et~al.}(2020){Dewangan}, {Ojha}, {Sharma}, {Palacio},
  {Bhadari}, \& {Das}}]{Dewangan2020}
{Dewangan}, L.~K., {Ojha}, D.~K., {Sharma}, S., {et~al.} 2020, \apj, 903, 13,
  \dodoi{10.3847/1538-4357/abb827}

\bibitem[{{Dewangan} {et~al.}(2017{\natexlab{c}}){Dewangan}, {Ojha}, \&
  {Zinchenko}}]{Dewangan17N49}
{Dewangan}, L.~K., {Ojha}, D.~K., \& {Zinchenko}, I. 2017{\natexlab{c}}, \apj,
  851, 140, \dodoi{10.3847/1538-4357/aa9be2}

\bibitem[{{Dewangan} {et~al.}(2019){Dewangan}, {Pirogov}, {Ryabukhina}, {Ojha},
  \& {Zinchenko}}]{Dewangan19S242}
{Dewangan}, L.~K., {Pirogov}, L.~E., {Ryabukhina}, O.~L., {Ojha}, D.~K., \&
  {Zinchenko}, I. 2019, \apj, 877, 1, \dodoi{10.3847/1538-4357/ab1aa6}

\bibitem[{{Dewangan} {et~al.}(2022){Dewangan}, {Zinchenko}, {Zemlyanukha},
  {Liu}, {Su}, {Kurtz}, {Ojha}, {Pazukhin}, \& {Mayya}}]{dewangan22}
{Dewangan}, L.~K., {Zinchenko}, I.~I., {Zemlyanukha}, P.~M., {et~al.} 2022,
  \apj, 925, 41, \dodoi{10.3847/1538-4357/ac36dd}

\bibitem[{{Fischera} \& {Martin}(2012)}]{Fischera2012}
{Fischera}, J., \& {Martin}, P.~G. 2012, \aap, 542, A77,
  \dodoi{10.1051/0004-6361/201218961}

\bibitem[{{Frerking} {et~al.}(1982){Frerking}, {Langer}, \&
  {Wilson}}]{Frekring1982}
{Frerking}, M.~A., {Langer}, W.~D., \& {Wilson}, R.~W. 1982, \apj, 262, 590,
  \dodoi{10.1086/160451}

\bibitem[{{Fukui} {et~al.}(2021){Fukui}, {Habe}, {Inoue}, {Enokiya}, \&
  {Tachihara}}]{fukui21}
{Fukui}, Y., {Habe}, A., {Inoue}, T., {Enokiya}, R., \& {Tachihara}, K. 2021,
  \pasj, 73, S1, \dodoi{10.1093/pasj/psaa103}

\bibitem[{{Garden} {et~al.}(1991){Garden}, {Hayashi}, {Gatley}, {Hasegawa}, \&
  {Kaifu}}]{Garden91}
{Garden}, R.~P., {Hayashi}, M., {Gatley}, I., {Hasegawa}, T., \& {Kaifu}, N.
  1991, \apj, 374, 540, \dodoi{10.1086/170143}

\bibitem[{{Getman} {et~al.}(2007){Getman}, {Feigelson}, {Garmire}, {Broos}, \&
  {Wang}}]{Getman07}
{Getman}, K.~V., {Feigelson}, E.~D., {Garmire}, G., {Broos}, P., \& {Wang}, J.
  2007, \apj, 654, 316, \dodoi{10.1086/509112}

\bibitem[{{Guieu} {et~al.}(2010){Guieu}, {Rebull}, {Stauffer}, {Vrba},
  {Noriega-Crespo}, {Spuck}, {Roelofsen Moody}, {Sepulveda}, {Weehler},
  {Maranto}, {Cole}, {Flagey}, {Laher}, {Penprase}, {Ramirez}, \&
  {Stolovy}}]{Guieu10}
{Guieu}, S., {Rebull}, L.~M., {Stauffer}, J.~R., {et~al.} 2010, \apj, 720, 46,
  \dodoi{10.1088/0004-637X/720/1/46}

\bibitem[{{Gutermuth} {et~al.}(2009){Gutermuth}, {Megeath}, {Myers}, {Allen},
  {Pipher}, \& {Fazio}}]{Gutermuth2009}
{Gutermuth}, R.~A., {Megeath}, S.~T., {Myers}, P.~C., {et~al.} 2009, \apjs,
  184, 18, \dodoi{10.1088/0067-0049/184/1/18}

\bibitem[{{Hacar} \& {Tafalla}(2011)}]{Hacar2011}
{Hacar}, A., \& {Tafalla}, M. 2011, \aap, 533, A34,
  \dodoi{10.1051/0004-6361/201117039}

\bibitem[{{Hacar} {et~al.}(2018){Hacar}, {Tafalla}, {Forbrich}, {Alves},
  {Meingast}, {Grossschedl}, \& {Teixeira}}]{Hacar2018}
{Hacar}, A., {Tafalla}, M., {Forbrich}, J., {et~al.} 2018, \aap, 610, A77,
  \dodoi{10.1051/0004-6361/201731894}

\bibitem[{{Hacar} {et~al.}(2013){Hacar}, {Tafalla}, {Kauffmann}, \&
  {Kov{\'a}cs}}]{Hacar2013}
{Hacar}, A., {Tafalla}, M., {Kauffmann}, J., \& {Kov{\'a}cs}, A. 2013, \aap,
  554, A55, \dodoi{10.1051/0004-6361/201220090}

\bibitem[{{Haemmerl{\'e}} {et~al.}(2016){Haemmerl{\'e}}, {Eggenberger},
  {Meynet}, {Maeder}, \& {Charbonnel}}]{Haemmerle2016}
{Haemmerl{\'e}}, L., {Eggenberger}, P., {Meynet}, G., {Maeder}, A., \&
  {Charbonnel}, C. 2016, \aap, 585, A65, \dodoi{10.1051/0004-6361/201527202}

\bibitem[{{Hartmann} {et~al.}(2005){Hartmann}, {Megeath}, {Allen}, {Luhman},
  {Calvet}, {D'Alessio}, {Franco-Hernandez}, \& {Fazio}}]{Hartmann05}
{Hartmann}, L., {Megeath}, S.~T., {Allen}, L., {et~al.} 2005, \apj, 629, 881,
  \dodoi{10.1086/431472}

\bibitem[{{Helfand} {et~al.}(2006){Helfand}, {Becker}, {White}, {Fallon}, \&
  {Tuttle}}]{Helfand06}
{Helfand}, D.~J., {Becker}, R.~H., {White}, R.~L., {Fallon}, A., \& {Tuttle},
  S. 2006, \aj, 131, 2525, \dodoi{10.1086/503253}

\bibitem[{{Hoemann} {et~al.}(2022){Hoemann}, {Heigl}, \&
  {Burkert}}]{Hoemann2022arXiv}
{Hoemann}, E., {Heigl}, S., \& {Burkert}, A. 2022, arXiv e-prints,
  arXiv:2203.07002.
\newblock \doarXiv{2203.07002}

\bibitem[{{Hunter} {et~al.}(1997){Hunter}, {Phillips}, \& {Menten}}]{Hunter97}
{Hunter}, T.~R., {Phillips}, T.~G., \& {Menten}, K.~M. 1997, \apj, 478, 283,
  \dodoi{10.1086/303775}

\bibitem[{{Inutsuka} \& {Miyama}(1992)}]{Inutsuka1992}
{Inutsuka}, S.-I., \& {Miyama}, S.~M. 1992, \apj, 388, 392,
  \dodoi{10.1086/171162}

\bibitem[{{Inutsuka} \& {Miyama}(1997)}]{Inutsuka1997}
{Inutsuka}, S.-i., \& {Miyama}, S.~M. 1997, \apj, 480, 681,
  \dodoi{10.1086/303982}

\bibitem[{{Kainulainen} {et~al.}(2016){Kainulainen}, {Hacar}, {Alves},
  {Beuther}, {Bouy}, \& {Tafalla}}]{Kainulainen2016}
{Kainulainen}, J., {Hacar}, A., {Alves}, J., {et~al.} 2016, \aap, 586, A27,
  \dodoi{10.1051/0004-6361/201526017}

\bibitem[{{Kirk} {et~al.}(2013){Kirk}, {Myers}, {Bourke}, {Gutermuth},
  {Hedden}, \& {Wilson}}]{Kirk2013}
{Kirk}, H., {Myers}, P.~C., {Bourke}, T.~L., {et~al.} 2013, \apj, 766, 115,
  \dodoi{10.1088/0004-637X/766/2/115}

\bibitem[{{Lawrence} {et~al.}(2007){Lawrence}, {Warren}, {Almaini}, {Edge},
  {Hambly}, {Jameson}, {Lucas}, {Casali}, {Adamson}, {Dye}, {Emerson},
  {Foucaud}, {Hewett}, {Hirst}, {Hodgkin}, {Irwin}, {Lodieu}, {McMahon},
  {Simpson}, {Smail}, {Mortlock}, \& {Folger}}]{Lawrence07}
{Lawrence}, A., {Warren}, S.~J., {Almaini}, O., {et~al.} 2007, \mnras, 379,
  1599, \dodoi{10.1111/j.1365-2966.2007.12040.x}

\bibitem[{{Li} {et~al.}(2013){Li}, {Wyrowski}, {Menten}, \&
  {Belloche}}]{Li2013}
{Li}, G.-X., {Wyrowski}, F., {Menten}, K., \& {Belloche}, A. 2013, \aap, 559,
  A34, \dodoi{10.1051/0004-6361/201322411}

\bibitem[{{Liu} {et~al.}(2012){Liu}, {Jim{\'e}nez-Serra}, {Ho}, {Chen},
  {Zhang}, \& {Li}}]{Liu2012}
{Liu}, H.~B., {Jim{\'e}nez-Serra}, I., {Ho}, P. T.~P., {et~al.} 2012, \apj,
  756, 10, \dodoi{10.1088/0004-637X/756/1/10}

\bibitem[{{Liu} {et~al.}(2019){Liu}, {Stutz}, \& {Yuan}}]{Liu19}
{Liu}, H.-L., {Stutz}, A., \& {Yuan}, J.-H. 2019, \mnras, 487, 1259,
  \dodoi{10.1093/mnras/stz1340}

\bibitem[{{Mangum} \& {Shirley}(2015)}]{Mangum15}
{Mangum}, J.~G., \& {Shirley}, Y.~L. 2015, \pasp, 127, 266,
  \dodoi{10.1086/680323}

\bibitem[{{McKee} \& {Ostriker}(2007)}]{McKee2007}
{McKee}, C.~F., \& {Ostriker}, E.~C. 2007, \araa, 45, 565,
  \dodoi{10.1146/annurev.astro.45.051806.110602}

\bibitem[{{Men'shchikov}(2021)}]{Men2021}
{Men'shchikov}, A. 2021, \aap, 649, A89, \dodoi{10.1051/0004-6361/202039913}

\bibitem[{{Molinari} {et~al.}(2010){Molinari}, {Swinyard}, {Bally}, {Barlow},
  {Bernard}, {Martin}, {Moore}, {Noriega-Crespo}, {Plume}, {Testi}, {Zavagno},
  {Abergel}, {Ali}, {Anderson}, {Andr{\'e}}, {Baluteau}, {Battersby},
  {Beltr{\'a}n}, {Benedettini}, {Billot}, {Blommaert}, {Bontemps}, {Boulanger},
  {Brand}, {Brunt}, {Burton}, {Calzoletti}, {Carey}, {Caselli}, {Cesaroni},
  {Cernicharo}, {Chakrabarti}, {Chrysostomou}, {Cohen}, {Compiegne}, {de
  Bernardis}, {de Gasperis}, {di Giorgio}, {Elia}, {Faustini}, {Flagey},
  {Fukui}, {Fuller}, {Ganga}, {Garcia-Lario}, {Glenn}, {Goldsmith}, {Griffin},
  {Hoare}, {Huang}, {Ikhenaode}, {Joblin}, {Joncas}, {Juvela}, {Kirk},
  {Lagache}, {Li}, {Lim}, {Lord}, {Marengo}, {Marshall}, {Masi}, {Massi},
  {Matsuura}, {Minier}, {Miville-Desch{\^e}nes}, {Montier}, {Morgan}, {Motte},
  {Mottram}, {M{\"u}ller}, {Natoli}, {Neves}, {Olmi}, {Paladini}, {Paradis},
  {Parsons}, {Peretto}, {Pestalozzi}, {Pezzuto}, {Piacentini}, {Piazzo},
  {Polychroni}, {Pomar{\`e}s}, {Popescu}, {Reach}, {Ristorcelli}, {Robitaille},
  {Robitaille}, {Rod{\'o}n}, {Roy}, {Royer}, {Russeil}, {Saraceno}, {Sauvage},
  {Schilke}, {Schisano}, {Schneider}, {Schuller}, {Schulz}, {Sibthorpe},
  {Smith}, {Smith}, {Spinoglio}, {Stamatellos}, {Strafella}, {Stringfellow},
  {Sturm}, {Taylor}, {Thompson}, {Traficante}, {Tuffs}, {Umana}, {Valenziano},
  {Vavrek}, {Veneziani}, {Viti}, {Waelkens}, {Ward-Thompson}, {White},
  {Wilcock}, {Wyrowski}, {Yorke}, \& {Zhang}}]{Molinari10}
{Molinari}, S., {Swinyard}, B., {Bally}, J., {et~al.} 2010, \aap, 518, L100,
  \dodoi{10.1051/0004-6361/201014659}

\bibitem[{{Motte} {et~al.}(2018){Motte}, {Bontemps}, \& {Louvet}}]{Motte2018}
{Motte}, F., {Bontemps}, S., \& {Louvet}, F. 2018, \araa, 56, 41,
  \dodoi{10.1146/annurev-astro-091916-055235}

\bibitem[{{Myers}(2009)}]{Myers09}
{Myers}, P.~C. 2009, \apj, 700, 1609, \dodoi{10.1088/0004-637X/700/2/1609}

\bibitem[{{Paron} {et~al.}(2009){Paron}, {Cichowolski}, \& {Ortega}}]{Paron09}
{Paron}, S., {Cichowolski}, S., \& {Ortega}, M.~E. 2009, \aap, 506, 789,
  \dodoi{10.1051/0004-6361/200912646}

\bibitem[{{Peretto} {et~al.}(2013){Peretto}, {Fuller}, {Duarte-Cabral},
  {Avison}, {Hennebelle}, {Pineda}, {Andr{\'e}}, {Bontemps}, {Motte},
  {Schneider}, \& {Molinari}}]{Peretto2013}
{Peretto}, N., {Fuller}, G.~A., {Duarte-Cabral}, A., {et~al.} 2013, \aap, 555,
  A112, \dodoi{10.1051/0004-6361/201321318}

\bibitem[{{Pon} {et~al.}(2012){Pon}, {Toal{\'a}}, {Johnstone},
  {V{\'a}zquez-Semadeni}, {Heitsch}, \& {G{\'o}mez}}]{Pon12}
{Pon}, A., {Toal{\'a}}, J.~A., {Johnstone}, D., {et~al.} 2012, \apj, 756, 145,
  \dodoi{10.1088/0004-637X/756/2/145}

\bibitem[{{Ragan} {et~al.}(2014){Ragan}, {Henning}, {Tackenberg}, {Beuther},
  {Johnston}, {Kainulainen}, \& {Linz}}]{Ragan2014}
{Ragan}, S.~E., {Henning}, T., {Tackenberg}, J., {et~al.} 2014, \aap, 568, A73,
  \dodoi{10.1051/0004-6361/201423401}

\bibitem[{{Rathborne} {et~al.}(2009){Rathborne}, {Johnson}, {Jackson}, {Shah},
  \& {Simon}}]{Rathborne09}
{Rathborne}, J.~M., {Johnson}, A.~M., {Jackson}, J.~M., {Shah}, R.~Y., \&
  {Simon}, R. 2009, \apjs, 182, 131, \dodoi{10.1088/0067-0049/182/1/131}

\bibitem[{{Rigby} {et~al.}(2016){Rigby}, {Moore}, {Plume}, {Eden}, {Urquhart},
  {Thompson}, {Mottram}, {Brunt}, {Butner}, {Dempsey}, {Gibson}, {Hatchell},
  {Jenness}, {Kuno}, {Longmore}, {Morgan}, {Polychroni}, {Thomas}, {White}, \&
  {Zhu}}]{Rigby16}
{Rigby}, A.~J., {Moore}, T.~J.~T., {Plume}, R., {et~al.} 2016, \mnras, 456,
  2885, \dodoi{10.1093/mnras/stv2808}

\bibitem[{{Rivera-Ingraham} {et~al.}(2010){Rivera-Ingraham}, {Ade}, {Bock},
  {Chapin}, {Devlin}, {Dicker}, {Griffin}, {Gundersen}, {Halpern}, {Hargrave},
  {Hughes}, {Klein}, {Marsden}, {Martin}, {Mauskopf}, {Netterfield}, {Olmi},
  {Patanchon}, {Rex}, {Scott}, {Semisch}, {Truch}, {Tucker}, {Tucker}, {Viero},
  \& {Wiebe}}]{Rivera-Ingraham10}
{Rivera-Ingraham}, A., {Ade}, P. A.~R., {Bock}, J.~J., {et~al.} 2010, \apj,
  723, 915, \dodoi{10.1088/0004-637X/723/1/915}

\bibitem[{{Schneider} {et~al.}(2012){Schneider}, {Csengeri}, {Hennemann},
  {Motte}, {Didelon}, {Federrath}, {Bontemps}, {Di Francesco}, {Arzoumanian},
  {Minier}, {Andr{\'e}}, {Hill}, {Zavagno}, {Nguyen-Luong}, {Attard},
  {Bernard}, {Elia}, {Fallscheer}, {Griffin}, {Kirk}, {Klessen}, {K{\"o}nyves},
  {Martin}, {Men'shchikov}, {Palmeirim}, {Peretto}, {Pestalozzi}, {Russeil},
  {Sadavoy}, {Sousbie}, {Testi}, {Tremblin}, {Ward-Thompson}, \&
  {White}}]{Schneider2012}
{Schneider}, N., {Csengeri}, T., {Hennemann}, M., {et~al.} 2012, \aap, 540,
  L11, \dodoi{10.1051/0004-6361/201118566}

\bibitem[{{Schuller} {et~al.}(2009){Schuller}, {Menten}, {Contreras},
  {Wyrowski}, {Schilke}, {Bronfman}, {Henning}, {Walmsley}, {Beuther},
  {Bontemps}, {Cesaroni}, {Deharveng}, {Garay}, {Herpin}, {Lefloch}, {Linz},
  {Mardones}, {Minier}, {Molinari}, {Motte}, {Nyman}, {Reveret}, {Risacher},
  {Russeil}, {Schneider}, {Testi}, {Troost}, {Vasyunina}, {Wienen}, {Zavagno},
  {Kovacs}, {Kreysa}, {Siringo}, \& {Wei{\ss}}}]{Schuller09}
{Schuller}, F., {Menten}, K.~M., {Contreras}, Y., {et~al.} 2009, \aap, 504,
  415, \dodoi{10.1051/0004-6361/200811568}

\bibitem[{{Skrutskie} {et~al.}(2006){Skrutskie}, {Cutri}, {Stiening},
  {Weinberg}, {Schneider}, {Carpenter}, {Beichman}, {Capps}, {Chester},
  {Elias}, {Huchra}, {Liebert}, {Lonsdale}, {Monet}, {Price}, {Seitzer},
  {Jarrett}, {Kirkpatrick}, {Gizis}, {Howard}, {Evans}, {Fowler}, {Fullmer},
  {Hurt}, {Light}, {Kopan}, {Marsh}, {McCallon}, {Tam}, {Van Dyk}, \&
  {Wheelock}}]{Skrutskie06}
{Skrutskie}, M.~F., {Cutri}, R.~M., {Stiening}, R., {et~al.} 2006, \aj, 131,
  1163, \dodoi{10.1086/498708}

\bibitem[{{Smith} {et~al.}(2016){Smith}, {Glover}, {Klessen}, \&
  {Fuller}}]{Smith2016}
{Smith}, R.~J., {Glover}, S. C.~O., {Klessen}, R.~S., \& {Fuller}, G.~A. 2016,
  \mnras, 455, 3640, \dodoi{10.1093/mnras/stv2559}

\bibitem[{{Solomon} {et~al.}(1987){Solomon}, {Rivolo}, {Barrett}, \&
  {Yahil}}]{Solomon1987}
{Solomon}, P.~M., {Rivolo}, A.~R., {Barrett}, J., \& {Yahil}, A. 1987, \apj,
  319, 730, \dodoi{10.1086/165493}

\bibitem[{{Su} {et~al.}(2015){Su}, {Zhang}, {Shao}, \& {Yang}}]{Su2015}
{Su}, Y., {Zhang}, S., {Shao}, X., \& {Yang}, J. 2015, \apj, 811, 134,
  \dodoi{10.1088/0004-637X/811/2/134}

\bibitem[{{Tafalla} \& {Hacar}(2015)}]{Tafalla2015}
{Tafalla}, M., \& {Hacar}, A. 2015, \aap, 574, A104,
  \dodoi{10.1051/0004-6361/201424576}

\bibitem[{{Tig{\'e}} {et~al.}(2017){Tig{\'e}}, {Motte}, {Russeil}, {Zavagno},
  {Hennemann}, {Schneider}, {Hill}, {Nguyen Luong}, {Di Francesco}, {Bontemps},
  {Louvet}, {Didelon}, {K{\"o}nyves}, {Andr{\'e}}, {Leuleu}, {Bardagi},
  {Anderson}, {Arzoumanian}, {Benedettini}, {Bernard}, {Elia}, {Figueira},
  {Kirk}, {Martin}, {Minier}, {Molinari}, {Nony}, {Persi}, {Pezzuto},
  {Polychroni}, {Rayner}, {Rivera-Ingraham}, {Roussel}, {Rygl}, {Spinoglio}, \&
  {White}}]{Tige2017}
{Tig{\'e}}, J., {Motte}, F., {Russeil}, D., {et~al.} 2017, \aap, 602, A77,
  \dodoi{10.1051/0004-6361/201628989}

\bibitem[{{Toal{\'a}} {et~al.}(2012){Toal{\'a}}, {V{\'a}zquez-Semadeni}, \&
  {G{\'o}mez}}]{Toala2012}
{Toal{\'a}}, J.~A., {V{\'a}zquez-Semadeni}, E., \& {G{\'o}mez}, G.~C. 2012,
  \apj, 744, 190, \dodoi{10.1088/0004-637X/744/2/190}

\bibitem[{{Trevi{\~n}o-Morales} {et~al.}(2019){Trevi{\~n}o-Morales}, {Fuente},
  {S{\'a}nchez-Monge}, {Kainulainen}, {Didelon}, {Suri}, {Schneider},
  {Ballesteros-Paredes}, {Lee}, {Hennebelle}, {Pilleri},
  {Gonz{\'a}lez-Garc{\'\i}a}, {Kramer}, {Garc{\'\i}a-Burillo}, {Luna},
  {Goicoechea}, {Tremblin}, \& {Geen}}]{Morales2019}
{Trevi{\~n}o-Morales}, S.~P., {Fuente}, A., {S{\'a}nchez-Monge}, {\'A}.,
  {et~al.} 2019, \aap, 629, A81, \dodoi{10.1051/0004-6361/201935260}

\bibitem[{{Umemoto} {et~al.}(2017){Umemoto}, {Minamidani}, {Kuno}, {Fujita},
  {Matsuo}, {Nishimura}, {Torii}, {Tosaki}, {Kohno}, {Kuriki}, {Tsuda},
  {Hirota}, {Ohashi}, {Yamagishi}, {Handa}, {Nakanishi}, {Omodaka}, {Koide},
  {Matsumoto}, {Onishi}, {Tokuda}, {Seta}, {Kobayashi}, {Tachihara}, {Sano},
  {Hattori}, {Onodera}, {Oasa}, {Kamegai}, {Tsuboi}, {Sofue}, {Higuchi},
  {Chibueze}, {Mizuno}, {Honma}, {Muller}, {Inoue}, {Morokuma-Matsui},
  {Shinnaga}, {Ozawa}, {Takahashi}, {Yoshiike}, {Costes}, \&
  {Kuwahara}}]{umemoto17}
{Umemoto}, T., {Minamidani}, T., {Kuno}, N., {et~al.} 2017, \pasj, 69, 78,
  \dodoi{10.1093/pasj/psx061}

\bibitem[{{Urquhart} {et~al.}(2018){Urquhart}, {K{\"o}nig}, {Giannetti},
  {Leurini}, {Moore}, {Eden}, {Pillai}, {Thompson}, {Braiding}, {Burton},
  {Csengeri}, {Dempsey}, {Figura}, {Froebrich}, {Menten}, {Schuller}, {Smith},
  \& {Wyrowski}}]{Urquhart18}
{Urquhart}, J.~S., {K{\"o}nig}, C., {Giannetti}, A., {et~al.} 2018, \mnras,
  473, 1059, \dodoi{10.1093/mnras/stx2258}

\bibitem[{{Vig} {et~al.}(2006){Vig}, {Ghosh}, {Kulkarni}, {Ojha}, \&
  {Verma}}]{Vig06}
{Vig}, S., {Ghosh}, S.~K., {Kulkarni}, V.~K., {Ojha}, D.~K., \& {Verma}, R.~P.
  2006, \apj, 637, 400, \dodoi{10.1086/498389}

\bibitem[{Virtanen {et~al.}(2020)Virtanen, Gommers, Oliphant, Haberland, Reddy,
  Cournapeau, Burovski, Peterson, Weckesser, Bright, {van der Walt}, Brett,
  Wilson, Millman, Mayorov, Nelson, Jones, Kern, Larson, Carey, Polat, Feng,
  Moore, {VanderPlas}, Laxalde, Perktold, Cimrman, Henriksen, Quintero, Harris,
  Archibald, Ribeiro, Pedregosa, {van Mulbregt}, \& {SciPy 1.0
  Contributors}}]{scipy20}
Virtanen, P., Gommers, R., Oliphant, T.~E., {et~al.} 2020, Nature Methods, 17,
  261, \dodoi{10.1038/s41592-019-0686-2}

\bibitem[{{Wang} {et~al.}(2020){Wang}, {Koch}, {Galv{\'a}n-Madrid}, {Lai},
  {Liu}, {Lin}, \& {Pattle}}]{Wang2020}
{Wang}, J.-W., {Koch}, P.~M., {Galv{\'a}n-Madrid}, R., {et~al.} 2020, \apj,
  905, 158, \dodoi{10.3847/1538-4357/abc74e}

\bibitem[{{Wang} {et~al.}(2016){Wang}, {Testi}, {Burkert}, {Walmsley},
  {Beuther}, \& {Henning}}]{Wang2016}
{Wang}, K., {Testi}, L., {Burkert}, A., {et~al.} 2016, \apjs, 226, 9,
  \dodoi{10.3847/0067-0049/226/1/9}

\bibitem[{{Wood} \& {Churchwell}(1989)}]{Wood89}
{Wood}, D. O.~S., \& {Churchwell}, E. 1989, \apjs, 69, 831,
  \dodoi{10.1086/191329}

\bibitem[{{Wright} {et~al.}(2010){Wright}, {Eisenhardt}, {Mainzer}, {Ressler},
  {Cutri}, {Jarrett}, {Kirkpatrick}, {Padgett}, {McMillan}, {Skrutskie},
  {Stanford}, {Cohen}, {Walker}, {Mather}, {Leisawitz}, {Gautier}, {McLean},
  {Benford}, {Lonsdale}, {Blain}, {Mendez}, {Irace}, {Duval}, {Liu}, {Royer},
  {Heinrichsen}, {Howard}, {Shannon}, {Kendall}, {Walsh}, {Larsen}, {Cardon},
  {Schick}, {Schwalm}, {Abid}, {Fabinsky}, {Naes}, \& {Tsai}}]{Wright10}
{Wright}, E.~L., {Eisenhardt}, P. R.~M., {Mainzer}, A.~K., {et~al.} 2010, \aj,
  140, 1868, \dodoi{10.1088/0004-6256/140/6/1868}

\bibitem[{{Xu} {et~al.}(2018){Xu}, {Xu}, {Zhang}, {Liu}, {Yu}, {Ning}, \&
  {Ju}}]{Xu18}
{Xu}, J.-L., {Xu}, Y., {Zhang}, C.-P., {et~al.} 2018, \aap, 609, A43,
  \dodoi{10.1051/0004-6361/201629189}

\bibitem[{{Yu} {et~al.}(2019){Yu}, {Xu}, \& {Wang}}]{Yu2019}
{Yu}, N.-P., {Xu}, J.-L., \& {Wang}, J.-J. 2019, \aap, 622, A155,
  \dodoi{10.1051/0004-6361/201832962}

\bibitem[{{Zernickel} {et~al.}(2013){Zernickel}, {Schilke}, \&
  {Smith}}]{Zernickel2013}
{Zernickel}, A., {Schilke}, P., \& {Smith}, R.~J. 2013, \aap, 554, L2,
  \dodoi{10.1051/0004-6361/201321425}

\bibitem[{{Zhang} {et~al.}(2019){Zhang}, {Kainulainen}, {Mattern}, {Fang}, \&
  {Henning}}]{Zhang19}
{Zhang}, M., {Kainulainen}, J., {Mattern}, M., {Fang}, M., \& {Henning}, T.
  2019, \aap, 622, A52, \dodoi{10.1051/0004-6361/201732400}

\end{thebibliography}
\bibliographystyle{aasjournal}



%
%
%

\end{document}